\documentclass[pdflatex,sn-mathphys]{sn-jnl}% Math and Physical Sciences Reference Style
\usepackage{bm}
\usepackage{cleveref}
\crefformat{equation}{(#2#1#3)}
\newcommand\norm[1]{\left\lVert#1\right\rVert}

\newcommand{\matr}[1]{#1}

\jyear{2023}%

\theoremstyle{thmstyleone}%
\theoremstyle{thmstyletwo}%
\newtheorem{remark}{Remark}%

\theoremstyle{thmstylethree}%

\begin{document}
\UseRawInputEncoding
\title[Article Title]{Parametric model order reduction for a wildland fire model via the shifted POD based deep learning method}

\author*[1,2]{\fnm{Shubhaditya} \sur{Burela}}\email{burela@tnt.tu-berlin.de}

\author[1,3]{\fnm{Philipp} \sur{Krah}}\email{philipp.krah@univ-amu.fr}

\author[1,2]{\fnm{Julius} \sur{Reiss}}\email{reiss@tnt.tu-berlin.de}

\affil*[1]{\orgdiv{Institute of Mathematics}, \orgname{Technische  Universit\"at Berlin}, \orgaddress{\street{Stra\ss e des 17. Juni 136}, \city{Berlin}, \postcode{10623}, \state{Berlin}, \country{Germany}}}
\affil[2]{\orgdiv{Institute of Fluid Mechanics and Technical Acoustics}, \orgname{Technische  Universit\"at Berlin}, \orgaddress{\street{M\"uller-Breslau-Stra\ss e 15}, \city{Berlin}, \postcode{10623}, \state{Berlin}, \country{Germany}}}
\affil[3]{\orgdiv{Institut de Math\'ematiques de Marseille (I2M)}, \orgname{Aix-Marseille Universit\'e}, \orgaddress{\street{39 Rue Joliot-Curie}, \city{Marseille}, \postcode{13453}, \country{France}}}

\abstract{Parametric model order reduction techniques often struggle to accurately represent transport-dominated phenomena due to a slowly decaying Kolmogorov \textit{n}-width. To address this challenge, we propose a non-intrusive, data-driven methodology that combines the shifted proper orthogonal decomposition (POD) with deep learning. Specifically, the shifted POD technique is utilized to derive a high-fidelity, low-dimensional model of the flow, which is subsequently utilized as input to a deep learning framework to forecast the flow dynamics under various temporal and parameter conditions.
The efficacy of the proposed approach is demonstrated through the analysis of one- and two-dimensional wildland fire models with varying reaction rates, and its performance is evaluated using multiple error measures. The results indicate that the proposed approach yields highly accurate results within the percent range, while also enabling rapid prediction of system states within seconds.}  



\maketitle

\section*{Acknowledgement}
We gratefully acknowledge the support of the Deutsche Forschungsgemeinschaft (DFG) as part of GRK2433 DAEDALUS.
The authors were granted access to the HPC resources of IDRIS under allocation No. AD012A01664R1 attributed by Grand \'Equipement National de Calcul Intensif (GENCI). 
Centre de Calcul Intensif d'Aix-Marseille is acknowledged for granting access to its high-performance computing resources. We also thank Tobias Breiten for his valuable comments and feedback.

\section{Introduction}\label{sec:introduction}
Forecasting the spread of forest fires has become an important aspect of civil protection \cite{european_commission_joint_research_centre_european_2020}. The intensification of extreme droughts and heat waves due to climate change has increased the frequency and severity of large forest fires worldwide. Consequently, predicting the risk and identifying the causes behind 
these events has become crucial in comprehending the connection between climate and land surface and helping manage forest fires \cite{kondylatos_wildfire_2022}. Numerous studies such as those referenced in \cite{sayad_predictive_2019, vilar_modelling_2021, valero_multifidelity_2021, mandel_wildland_2008} have been undertaken to model and forecast forest fires utilizing wildland fire models that can be simulated under various scenarios. Through this paper, we present a novel approach for predicting the spread of forest fires through a combination of dimensionality reduction and deep learning (DL).  

The phenomena of wildland fires are usually governed by parametric non-linear time-dependent partial differential equations (PDEs) with different scenarios being achieved by changing physical parameters in the system of equations. However, solving such equations for modeling large-scale wildland fires for the numerous parameters is unaffordable in real-time emergency situations requiring significant computing power and time. To speed up the numerical simulations  reduced-order modeling provides a promising strategy.

Reduced order modeling (ROM) for the parameterized PDEs usually relies on an offline-online computational splitting \cite{benner_model_2017}. The expensive task of building the low-dimensional subspace from the full order model (FOM) snapshots is performed once in the so-called offline stage, and the reduced order model (ROM) approximation corresponding to any new parameter value is computed in the so-called online stage. Classical projection-based methods such as proper orthogonal decomposition (POD) perform the dimensionality reduction by constructing a low-dimensional trial subspace by the leading modes of a singular value decomposition (SVD) and then projecting the system dynamics onto the constructed low-dimensional subspace using Petrov-Galerkin methods \cite{hesthaven_certified_2016, quarteroni_reduced_2016}. In many applications, namely with faster decay of the Kolmogorov $n$-width \cite{greif_decay_2019} (in turn the Hankel singular values \cite{unger_kolmogorov_2019}), a low-dimensional subspace with low approximation error if found with conventional methods like POD. But the problem arises when the dimension of the trial subspace needs to be large for a desired approximation quality. This is almost always the case with transport-dominated fluid systems (TDFS) like propagating flame fronts and traveling acoustic or shock waves \cite{krah_front_2022, huang_challenges_2018, ohlberger_nonlinear_2013}. 

To overcome the problem of slowly decaying singular values, model order reduction (MOR) methods for TDFS have emerged over the last couple of years. The work of \cite{rim_transport_2018} uses transport reversal where they do template fitting by posing a minimization problem. Along similar lines \cite{cagniart_model_2019} tries to transform/twist the set of solutions so that the combination of a proper shift and appropriate linear combination recovers accurate approximation. Similar approaches are followed in \cite{welper_transformed_2020, black_projection-based_2020}. Transport maps were explored in \cite{nonino_overcoming_2019}, approximating the field variable with a front shape function and a level set function for efficient model reduction is explored in \cite{krah_model_2021,krah_front_2022}. In general, most of these methods rely on an offline-online splitting where the computationally expensive offline step comprises a non-linear projection framework to capture the high-dimensional space and the online step constructs the ROM approximation intrusively for any unseen parameter value. Intrusive parameter predictions have already been utilized for the wildland fire model in \cite{black_efficient_2021}, where the shifted proper orthogonal decomposition (sPOD) \cite{reiss_shifted_2018} is combined with residual minimization \cite{lee_model_2020, black_projection-based_2020} to obtain an efficient ROM. More powerful formulations of the sPOD are discussed in \cite{reiss_optimization-based_2021, krah_non-linear_nodate}.
The sPOD has also been  used in \cite{mendible_dimensionality_2020} for predicting the detonation rotation waves using UnTWIST algorithm and in \cite{papapicco_neural_2022} for the prediction of shifts in nonlinear hyperbolic equations. 

However, the non-linearity of the dimensional reduction for transport phenomena  hinders the projection of even a linear PDE on a small basis, and thus it needs to be evaluated in the original dimension of the FOM in the online stage. This prevents any significant speedup during the simulation. In order to avoid scaling with the FOM dimension hyper-reduction techniques \cite{sarna_hyper-reduction_2021, carlberg_gnat_2013, barrault_empirical_2004} have been developed, one such being discrete empirical interpolation method (DEIM) \cite{jain_hyper-reduction_2017, chaturantabut_nonlinear_2010} which approximates non-linear terms by evaluating them at a few, carefully selected interpolation points and approximating all other components via interpolation in a low-dimensional space. This method also serves as the starting point for further extension of the method called shifted-DEIM by \cite{black_efficient_2021} for treating the wildland fire model. 

Another class of methods called online adaptive basis methods \cite{peherstorfer_model_2020, koch_dynamical_2007, etter_online_2020, dihlmann_model_2011} have also been introduced to circumvent the problem of slow-decaying Kolmogorov \textit{n}-width by exploiting the time-local low-rank structure of the TDFS. One such method called adaptive bases and adaptive sampling discrete empirical interpolation method (AADEIM) was introduced by Peherstorfer et.al (2020) in \cite{peherstorfer_model_2020}. It builds upon the argument that the solutions of TDFS are typically low-dimensional if considered locally in time. This notion is then exploited by approximating the FOM solution in local low-dimensional spaces that are adapted via a low-rank basis updates over time \cite{zimmermann_geometric_2018, peherstorfer_online_2015}. However, the problem with these adaptive versions is that they usually do not deliver speedups comparable to a traditional POD-DEIM approach, as shown for shallow water flows in \cite{koellermeier_macro-micro_2023}. To circumvent these problems, various non-intrusive methods have emerged in the recent past. These methods rely in general on projecting the high-dimensional dynamics onto a low-dimensional subspace usually by using the POD and then employing deep learning (DL) framework for modeling the reduced dynamics. 

A few attempts have been made to construct a low-dimensional subspace with artificial neural networks (ANNs) \cite{gonzalez_deep_2018, lee_model_2020}, shallow masked encoder \cite{kim_fast_2022} and methods like kernel POD \cite{salvador_non_2021} as well which allow for a non-linear representation of the projected high-dimensional dynamics. Subsequently DL techniques such as feed-forward neural network (FNN) based regression model \cite{san_neural_2018, salvador_non_2021, hesthaven_non-intrusive_2018, wang_non-intrusive_2019, berzins_standardized_2021} and Gaussian process regression \cite{kast_non-intrusive_2020, guo_data-driven_2019} have been employed to model the reduced dynamics. Model order reduction techniques based on a deep convolutional autoencoder are proposed in \cite{gonzalez_deep_2018, bhattacharya_model_2021}. Other DL techniques in the literature include long short-term memory (LSTMs) \cite{mohan_deep_2018} and recurrent neural networks (RNNs) \cite{j_nagoor_kani_reduced-order_2019}. 

In our research, we expand upon the non-intrusive parameter prediction framework introduced by Fresca et al. \cite{fresca_comprehensive_2021, fresca_pod-dl-rom_2022}. This involves constructing a trial subspace using Proper Orthogonal Decomposition (POD) and modeling the reduced dynamics using a convolutional autoencoder coupled with a Feedforward Neural Network (FNN). However, instead of the POD, we employ the sPOD for dimensionality reduction, as it avoids the issue of slowly decaying singular values.
On this base, we employ a deep FNN to learn the time and parameter-dependent amplitudes and shift transformations. This allows us to efficiently predict the states of the system for unseen parameter values.
A similar technique has been presented in Kovarnova et al. \cite{kovarnova_shifted_2022}, where they only predict in time. We apply our proposed approach to one- and two-dimensional wildland fire models, and present computational results to demonstrate its effectiveness.

The structure of the paper is as follows. In \Cref{sec:theory} we lay out the theoretical foundation concerning the mode decomposition methods, namely POD and sPOD along with the formulation describing the data-driven approach. A brief definition of the errors considered in our studies for model performance evaluation is also mentioned in \Cref{sec:theory}. In \Cref{sec:results} we present the numerical results for all the example test cases along with a timing study. Finally, the conclusions are drawn in \Cref{sec:conclusion}, and auxiliary tests are provided in the Appendix.

\section{Theory and background}\label{sec:theory}
In this section, we introduce the theoretical aspects of the presented methods. We study the relevant mode decomposition approaches followed by the data-driven approach. Furthermore, we scrutinize the types of errors that may arise from the implementation of the proposed method.
\subsection{Mode decomposition}
We specifically consider POD and sPOD for our analysis.
\subsubsection{Proper Orthogonal Decomposition (POD)}
POD is a method to extract optimal basis sets from a collection of snapshots. The snapshot matrix $Q \in \mathbb{R}^{M \times N}$ comprises of snapshots $q(\bm{x}_i, t_j, \bm{\mu}_j)$ arranged in a column-wise fashion for each time step $t_j$ where $j=1, \ldots, N$, $\bm{x}_i \in \Omega$ are the spatial grid points inside the domain $\Omega$ and $\bm{\mu}_j$ being the parameter dependency. The POD method approximates with the help of SVD the snapshot matrix $Q$ 
\begin{equation}\label{eq-def:POD}
    Q \approx \tilde{Q} = U_r \Sigma_r (V_r)^\top\, .
\end{equation}
Here, $r\ll \min(N, M)$ is the truncation rank, $\matr{\Sigma}_{r}=\mathrm{diag}{(\sigma_1,\dots,\sigma_{r})}$ is a diagonal matrix containing the singular values $\sigma_1\ge\sigma_2\ge\dots\sigma_{r}$ and $\matr{U}_{r}\in\mathbb{R}^{M\times {r}}$, $\matr{V}_{r}\in\mathbb{R}^{N\times {r}}$ are orthogonal matrices containing the left and right singular vectors. The reconstruction is optimal in the sense that the time-averaged least-square error of the POD approximation
\begin{equation}\label{eq-def:PODerr}
    \epsilon_r = \left\|Q - \tilde{Q}\right\|^2_{\mathrm{F}} = \sum^{\min(N, M)}_{k=r+1}\sigma_k
\end{equation}
is minimized.

\subsubsection{Shifted proper orthogonal decomposition (sPOD)}
The sPOD was introduced in \cite{reiss_shifted_2018} and further algorithmic developments were presented in \cite{reiss_optimization-based_2021, krah_non-linear_nodate} based on the optimization of singular values and in \cite{black_projection-based_2020, black_efficient_2021} based on the optimization of dyadic pairs and their shifts.
The sPOD method aims to decompose the snapshot matrix ${Q}=[q(\bm{x}_i,t_j, \bm{\mu}_j)]_{ij}\in\mathbb{R}^{M\times N}$ into multiple co-moving fields $\{\matr{Q}^k\in\mathbb{R}^{M\times N}\}_{k=1,\dots,f}$ (i.e.~$\{q^k(\bm{x}_i,t_j, \bm{\mu}_j)\}_{k=1,\dots,f}$). The decomposition follows as:
\begin{align}
    q(\bm{x},t,\bm{\mu}) &\approx   {\tilde{q}(\bm{x},t,\bm{\mu})} :=\sum_{k=1}^f \mathcal{T}^{{\Delta}^k} q^k(\bm{x},t,\bm{\mu})\,, \quad &\text{(continuous)}\\
    \matr{Q} &\approx {\tilde{\matr{Q}}} :=\sum_{k=1}^f T^{\bm{\Delta}^k} \matr{Q}^k\,, \quad &\text{(discrete)}
    \label{eq-def:sPOD-decomposition-discrete}
\end{align}
The interpolation based discrete transformation operators are given by $\{T^{\bm{\Delta^k}}\}_{k=1,\dots,f}$ where $\bm{\Delta}^k(t, \mu)={(\Delta^k(\bm{x}_1,t,\bm{\mu}),\dots,\Delta^k(\bm{x}_M,t,\bm{\mu}))}$ is the time dependent shift such that $T^{\pm\bm{\Delta}^k}(Q)_{ij} = q(\bm{x}_i \mp  {\Delta}^{k}(\bm{x}_i,t, \bm{\mu})_i, t_j, \bm{\mu}_j)$, and $\tilde{Q}$ is the approximate reconstruction of $Q$. The assumption is that for traveling wave systems the superposition \Cref{eq-def:sPOD-decomposition-discrete} can decompose the data more efficiently than the POD. This is because the traveling wave is only changing slowly in the co-moving data frame and can therefore be well approximated with only a few modes calculated with a truncated SVD:
\begin{align}
         \matr{Q}^k \approx \matr{U}_{r^k}^k\matr{\Sigma}_{r^k}^k (\matr{V}_{r^k}^k)^\top\,\quad k=1,\dots,f\,.
\label{eq-def:sPOD-Qk}
\end{align}
Here,  $r_k\ll N$ is the truncation rank of each co-moving field, $\matr{\Sigma}^k_{r_k}=\mathrm{diag}{(\sigma_1^k,\dots,\sigma_{r_k}^k)}$ is a diagonal matrix containing the singular values $\sigma_1^k\ge\sigma_2^k\ge\dots\sigma_{r_k}^k$ and $\matr{U}_{r_k}^k\in\mathbb{R}^{M\times {r_k}}$, $\matr{V}_{r_k}^k\in\mathbb{R}^{N\times {r_k}}$ are orthogonal matrices containing the left and right singular vectors. In this work we use the sPOD algorithm based on the minimization of the nuclear-one norm, which is presented in \cite{krah_non-linear_nodate} algorithm 8, and the corresponding tuning parameters for the algorithm are stated in the \Cref{apx:sPOD_parameters}.
\begin{remark}
In order to be more efficient than the POD, when disregarding additional degrees of freedom introduced by shifts, we assume that for the total number of degrees of freedom $r=\sum_{k=1}^f r_k$ the truncation error of the sPOD is smaller than the POD truncation error \cref{eq-def:PODerr}.
\end{remark}

\subsection{Data-driven approach}
We consider a non-linear parameterized dynamical system which often stems from a discretized PDE system like:%the wildland fire model:
\begin{equation}\label{eq_def:NPDS}
\begin{aligned}
    \bm{\dot{q}}(t,\bm{\mu}) &= \bm{f}(t, \bm{q}(t,\bm{\mu}),\mu) \quad t \in [0, T] \\
    \bm{q}(0,\bm{\mu}) &= \bm{q_0}(\bm{\mu})
\end{aligned}
\end{equation}
where, $\bm{q} \in \mathbb{R}^M$ is the parameterized solution to the problem, $\bm{q_0}$ is the initial data, $\bm{f}$ is the non-linear function describing the system dynamics and $\bm{\mu}$ is the parameter vector.
For simplicity we only consider a single parameter $\mu$ for our study, however, real-world scenarios may depend upon multiple parameters.
%In traditional projection-based ROMs for e.g. in POD-Galerkin ROMs, the trajectories of the solutions computed using the high-dimensional models are sought in the low-dimensional subspace $Col(U_r)$ of dimension $r \ll M$ spanned by $r$ columns of the subspace $U_r \in \mathbb{R}^{M \times r}$. Thus, the state is approximated as:
%\begin{equation}\label{eq:PODGalappx}
%    \bm{q}(t, \mu) = \bm{\tilde{q}}(t, \mu) = U_r \bm{q}_r(t,\mu).
%\end{equation}
%To model the reduced dynamics of the system, POD-Galerkin ROMs are widely used. 

%Although widely popular, the problem with these methods arises when the dimension $r$ of the low-dimensional subspace becomes large. This is usually the case with transport-dominated problems where the necessary dimension of the approximation subspace needs to be large in order for the projection error to be sufficiently low. These methods also rely on hyper-reduction strategies to obtain speedups for non-linear equations.

\subsubsection{Non-intrusive predictions with POD and sPOD}
%To circumvent the slow decay of singular values in transport-dominated systems, non-intrusive approaches based on deep learning framework are often employed for constructing efficient ROMs for parameter-dependent PDEs. In our work
To develop the new method, we closely follow the POD-DL-ROM method described by Fresca et. al %in their work
\cite{fresca_pod-dl-rom_2022}. The said method first builds a low-dimensional subspace $Col(U_r)$ which is spanned by the first $r$ singular vectors of the parameter snapshot matrix $Q$:
\begin{equation}\label{eq:par_snap_matr}
    Q = 
    \left[
      \begin{array}{ccccccc}
        \vert &  & \vert &  & \vert  &   & \vert \\
        \bm{q}(t_1,\mu_1)    & \ldots   &  \bm{q}(t_{N_t},\mu_1)  & \ldots  &  \bm{q}(t_1,\mu_{N_{p}})  &  \ldots & \bm{q}(t_{N_t},\mu_{N_{p}})  \\
        \vert &  & \vert &  & \vert  &   & \vert 
      \end{array}
    \right] \in \mathbb{R}^{M \times N}. 
\end{equation}
The matrix $Q$ is the collection of $N=N_p N_t$ number of FOM snapshots computed for different parameter instances $\mu_1, \ldots, \mu_{N_p}$ that are sampled over different time instances $t_1, \ldots, t_{N_t}$. Using POD on $Q$ and truncating to a rank $r \ll M$, we have
\begin{equation*}
    \tilde{Q} = U_r\Sigma_r (V_r)^\top\ \quad \text{where}, \quad A = \Sigma_r (V_r)^\top\
\end{equation*}
and
\begin{equation}\label{eq:par_ta_matr}
    A = 
    \left[
      \begin{array}{ccccccc}
        \vert & \vert &  &  & \vert\\
        \bm{a}(t_1,\mu_1) & \bm{a}(t_{2},\mu_1)  & \ldots  &  \ldots & \bm{a}(t_{N_t},\mu_{N_{p}})  \\
        \vert & \vert &  &  & \vert 
      \end{array}
    \right] \in \mathbb{R}^{r \times N}.
\end{equation}
A deep learning model is used to approximate the mapping $(t_1,\mu_1)\to \bm{a} $, by using $A$ as training data along with $P$:
\begin{equation}\label{eq:par_matr}
    P = 
    \left[
      \begin{array}{ccccccc}
        (t_1,\mu_1)    & \ldots   &  (t_{N_t},\mu_1)  & \ldots  &  (t_1,\mu_{N_{p}})  &  \ldots & (t_{N_t},\mu_{N_{p}}) 
      \end{array}
    \right] \in \mathbb{R}^{(n_\mu + 1)\times N}\,.
\end{equation}
The matrix $P$ corresponds to $n_{\mu}$ number of parameters along with the contribution of time $t$ as an additional parameter thus making $P\in \mathbb{R}^{(n_\mu + 1)\times N}$. 
By this, the network learns the mapping from parameter and time to amplitude, which is the mapping traditionally constructed by the Galerkin approach. Subsequently, once the network is trained, the prediction step is performed in order to generate the time amplitude for any $(t, \mu)$
\begin{align}
    \bm{a}(t, \mu) =
\left[
  \begin{array}{c}
    a_{1}(t, \mu)\\
    a_{2}(t, \mu)\\
    \vdots  \\
    a_{r}(t, \mu)
  \end{array}
\right] \approx \mathbf{N_{\mathrm{POD}}}(t, \mu) 
\end{align}
and the state $\bm{\tilde{q}'}$ for those unseen parameters is reconstructed as:
\begin{equation}
    \bm{\tilde{q}'}(t, \mu) = U_r \bm{a}(t, \mu)\,.
\end{equation}
More in-depth analysis of the POD-DL-ROM method along with the network architectures, working examples, and the error study can be found in \cite{fresca_comprehensive_2021, fresca_pod-dl-rom_2022}.

Since the  POD  is not suitable for transport-dominated systems, we present a novel non-intrusive model order reduction technique named sPOD-NN that uses sPOD for constructing the desired low-dimensional subspace. The sPOD is applied on the parameter snapshot matrix $Q$ from \Cref{eq:par_snap_matr}.
The sPOD decomposes the snapshot matrix $Q$ into co-moving frames $Q^k$. The algorithm also outputs the basis vectors $U^k_{r^k}$ for every frame with which the time amplitude matrix $A^k$ is extracted for all $k=1,\ldots,f$. Following \Cref{eq-def:sPOD-Qk} the time amplitudes are defined, 
\begin{equation}\label{eq:sPOD_extract}
    A^k = (U^k_{r^k})^\top\ Q^k \in \mathbb{R}^{r^k \times N},
\end{equation}
where
\begin{equation}\label{eq:par_ta_matr_sPOD}
    A^k = 
    \left[
      \begin{array}{ccccccc}
        \vert & \vert &  &  & \vert\\
        \bm{a}^k(t_1,\mu_1) & \bm{a}^k(t_{2},\mu_1)  & \ldots  &  \ldots & \bm{a}^k(t_{N_t},\mu_{N_{p}})  \\
        \vert & \vert &  &  & \vert 
      \end{array}
    \right].
\end{equation}
We now have the time amplitude matrices $A^k$ available for all $k$ along with the shifts. However, we note that the shifts for any new unseen parameter set $(t, \mu)$ are not available upfront. This poses a challenge because, in due course of reconstructing the final state for unseen parameter values, the shifts must be computed first.

\newcommand{\shiftDim}{n_\Delta}
\newcommand{\shiftMat}{S}
\paragraph{Shifts:}

The shifts $\bm{\Delta}^k(t,\mu) =(\Delta^k(\bm x_1,t,\mu),\dots,\Delta^k(\bm x_M,t,\mu)) \in \mathbb{R}^M$, used to encode the transport, are part of the description. 
They are in general dependent on the model parameters $\mu$, time $t$ and space $\bm x$. The flame propagation speed changes with the Arrhenius factor for the wildland fire model, and it needs to be predicted by the neural network. To have a low-dimensional description, the description of the shifts needs to be low-dimensional as well.        
We basically study two possible scenarios, where the first can be seen as a special case of the second:
\indent
\begin{minipage}{\dimexpr\textwidth-0.6cm}
    \paragraph{(a) Low-dimensional shifts:} 
    We look at a case where the shifts are independent of the spatial coordinates $\bm{x}\in\mathbb{R}^d$:
    \begin{equation}\label{eq:shifts_constant}
        {\Delta}^k(\bm x,t,\mu) = \sum_{i=1}^{\shiftDim^k}\bm e_i \:\underline{\Delta}^k_i(t,\mu) \,,\quad \text{where}\quad \shiftDim^k\le d\,.
    \end{equation}
    Here, $\bm e_i$ denotes the $i$th standard basis vector.
    For our one- and two-dimensional examples $d=1,2$. Since $\shiftDim^k\le d$ the coefficient vectors of the shifts  $\underline{\bm{\Delta}}^k(t,\mu)=(\underline{\Delta}_1^k(t,\mu),\dots,\underline{{\Delta}}^k_{\shiftDim^k}(t,\mu))$ are already low-dimensional and only depend on time and parameter. 
\end{minipage}
\indent
\begin{minipage}{\dimexpr\textwidth-0.6cm}
    \paragraph{(b) Low-rank description of high-dimensional shifts:} 
    The sPOD usually assumes a low-dimensional description of the shifts. However, for complicated systems, the shifts might depend on the spatial position $\bm{x}$ itself and are thus high-dimensional. Nevertheless, the shifts often have a low-rank structure and we thus assume, they can be well represented with the help of the POD:
    \begin{equation}
    \label{eq:shifts_pod}
        {\Delta}^k(\bm x,t,\mu)  \approx \sum_{n=1}^{\shiftDim^k} \Upsilon_n^k(\bm x) \underline{\Delta}_n^k(t,\mu)\,, \quad\text{with}\quad\shiftDim^k\ll N\,.
    \end{equation}
     Thus after we assemble the shift matrix $[\Delta^k(\bm{x}_i, t_j, \mu_j)]_{ij}\in\mathbb{R}^{M\times N}$ using a threshold algorithm (see \cref{sec:2dwithwind}), we decompose it with the help of the truncated SVD and obtain the low-dimensional shifts $\underline{\bm{\Delta}}^k(t,\mu)=(\underline{\Delta}_1^k(t,\mu),\dots,\underline{{\Delta}}^k_{\shiftDim^k}(t,\mu))$.
\end{minipage}
Once we obtain the $\underline{\bm{\Delta}}^k(t,\mu)$ we construct the training data by stacking all $A^k$ and $\underline{\bm{\Delta}}^k(t,\mu)$ to construct an $\hat{A} \in \mathbb{R}^{\left(\sum_k r^k + \sum_k \shiftDim^k\right) \times N}$ matrix as shown:
\begin{equation}\label{eq:par_ta_matr_sPOD_del}
    \hat{A} = 
    \left[
      \begin{array}{ccccc}
        \bm{a}^1(t_1,\mu_1) & \bm{a}^1(t_{2},\mu_1)  & \ldots  &  \ldots & \bm{a}^1(t_{N_t},\mu_{N_{p}})  \\
        \vdots & \vdots &  &  & \vdots \\
        \bm{a}^f(t_1,\mu_1) & \bm{a}^f(t_{2},\mu_1)  & \ldots  &  \ldots & \bm{a}^f(t_{N_t},\mu_{N_{p}})  \\
        \underline{\bm{\Delta}}^1(t_1,\mu_1) & \underline{\bm{\Delta}}^1(t_{2},\mu_1)  & \ldots  &  \ldots & \underline{\bm{\Delta}}^1(t_{N_t},\mu_{N_{p}})  \\
        \vdots & \vdots &  &  & \vdots \\ 
        \underline{\bm{\Delta}}^f(t_1,\mu_1) & \underline{\bm{\Delta}}^f(t_{2},\mu_1)  & \ldots  &  \ldots & \underline{\bm{\Delta}}^f(t_{N_t},\mu_{N_{p}})  \\
      \end{array}
    \right]\,.
\end{equation}
As for training the deep learning model, we consider $\hat{A}$ and the entries of $P$ matrix from \Cref{eq:par_matr} as the training data. After the successful training of the network, the time amplitudes and the shifts are predicted for $(t, \mu)$
    \begin{align}
        \bm{\hat{a}}(t, \mu) =
    \left[
      \begin{array}{c}
        \bm{a}^1(t, \mu)\\
        \vdots \\
        \bm{a}^f(t, \mu) \\
        \underline{\bm{\Delta}}^1(t, \mu)\\
        \vdots  \\
        \underline{\bm{\Delta}}^f(t, \mu)
      \end{array}
    \right] \approx \mathbf{N_{\mathrm{sPOD}}}(t, \mu)\,.
\end{align}
By reconstructing the shifts using \Cref{eq:shifts_constant} or \cref{eq:shifts_pod}  depending on the problem at hand the state $\bm{\tilde{q}'}$ for the unseen parameters can be reconstructed as
\begin{equation}
\label{eq:sPOD_state}
    \bm{\tilde{q}'}(t, \mu) = \sum^{f}_{k=1} T^{\bm{\Delta}^k(t, \mu)} \cdot \left(U^k_{r^k} \bm{a}^k(t, \mu)\right)\,.
\end{equation}
For later comparison studies, we define the degrees of freedom ($n_{\mathrm{dof}}$) as:
\begin{equation}\label{eq:dof}
    n_{\mathrm{dof}} = 
    \begin{cases}
        r, & \text{for POD-NN}\\
        \sum_k r^k + \sum_k \shiftDim^k,& \text{for sPOD-NN}
    \end{cases}
\end{equation}

\subsubsection{Interpolation method with sPOD}
For comparison we introduce another approach based on sPOD which we call sPOD-I (Interpolation) which instead of using the DL techniques uses an interpolation-based method in the online phase. The sPOD-I method first extracts the $P$ and $\hat{A}$ matrices as explained in \Cref{eq:par_matr} and \Cref{eq:par_ta_matr_sPOD_del} respectively. Subsequently, it then uses the scipy \cite{virtanen_scipy_2020} function   {\fontfamily{cmtt}\selectfont scipy.interpolate.griddata()} for interpolation. This function triangulates the input domain \cite{orourke_computational_1998} and performs barycentric interpolation on each triangle to construct an interpolant. For an interpolation point $(t, \mu)$ lying inside a triangle the interpolated value is given as
\begin{equation}\label{eq:sPOD-I}
    \bm{\hat{a}}(t, \mu) = \sum^{3}_{i=1} \alpha_i \bm{\hat{a}}(t_i, \mu_i), 
\end{equation}
 where $\alpha_i$ are barycentric coordinates with $\sum \alpha_i = 1$. Once the time amplitudes and the shifts are obtained, the final state can be reconstructed as shown in \Cref{eq:sPOD_state}.

\subsubsection{Network architecture}
We employ a deep FNN for the predictions. We use PyTorch \cite{paszke_pytorch_2019} for constructing the neural networks. For the architecture, we have an input layer, three hidden layers, and an output layer. The parameters of the network, the number of inputs and outputs which necessarily are the neurons for each layer are shown in \Cref{tab:app_network_params}. In the table, we have $p$ as the input for the input layer of the network where $p=n_{\mu} + 1$ as described in \Cref{eq:par_matr}. For our examples, we only consider a single parameter $\mu$ thus $p=2$. The number of outputs $u$ in the output layer is problem dependent and has been pointed out in the respective numerical examples. 
\begin{table}[h!]
\begin{center}
\begin{minipage}{\textwidth}
\caption{Deep feed-forward neural network parameters}\label{tab:app_network_params}
\centering
\begin{tabular}{||c c c c||} 
 \hline
 Layer & Num. of Inputs & Num. of Outputs & Activation function \\ [0.5ex] 
 \hline\hline
 Input layer & $p$ & 25 & ELU \\ 
 \hline
 $1^{st}$ hidden layer & 25 & 50 & ELU \\
 \hline
 $2^{nd}$ hidden layer & 50 & 75 & ELU \\
 \hline
 $3^{rd}$ hidden layer & 75 & 50 & LeakyReLU \\
 \hline
 Output layer & 50 & $u$ & - \\ [1ex] 
 \hline
\end{tabular}
\end{minipage}
\end{center}
\end{table}
We basically rely on the L1 loss function, namely MAE (Mean Absolute Error):
\begin{equation}
    \mathcal{L}_{\mathrm{MAE}} = 
    \begin{cases}
        \frac{1}{n_{\mathrm{dof}}}\left\|\bm{a}(t, \mu) - {\textbf{N}_{\mathrm{POD}}}(t, \mu)\right\|_1, & \text{for POD-NN}\\
        \frac{1}{n_{\mathrm{dof}}}\left\|\bm{\hat{a}}(t, \mu) - {\textbf{N}_{\mathrm{sPOD}}}(t, \mu)\right\|_{1},& \text{for sPOD-NN},
    \end{cases}
\end{equation}
where $n_{\mathrm{dof}}$ is defined in \cref{eq:dof}.
For improving the prediction accuracy we use data scaling. In this work, we use Min-Max scaling to scale our training data to the interval $[0, 1]$. Let us consider $\hat{A}$ from \Cref{eq:par_ta_matr_sPOD_del} and $P$ from \Cref{eq:par_matr} which serve as the training data for our network. The scaling for the matrix $P$ is given as:
\begin{equation}
    P^{\mathrm{scaled}}_{ij} = \frac{P_{ij} - \underset{j=1,\ldots,N_s}{\min}(P_{ij})}{\underset{j=1,\ldots,N_s}{\max}(P_{ij}) - \underset{j=1,\ldots,N_s}{\min}(P_{ij})}, \quad \text{for} \quad i=1, \ldots, n_{\mu} + 1
\end{equation}
whereas, for the matrix $\hat{A}$ where the time amplitudes and the shifts are stacked together, the scaling is performed separately for both quantities as shown here:
\begin{align}
 \hat{A}^{\mathrm{scaled}} =
\left[
  \begin{array}{c}
    \frac{\bm{A}_{ij} - \underset{i,j}{\min}(\bm{A}_{ij})}{\underset{i,j}{\max}(\bm{A}_{ij}) - \underset{i,j}{\min}(\bm{A}_{ij})}\\
    \vspace{0.05cm}\\
    \frac{\underline{\bm{\Delta}}_{ij} - \underset{i,j}{\min}(\underline{\bm{\Delta}}_{ij})}{\underset{i,j}{\max}(\underline{\bm{\Delta}}_{ij}) - \underset{i,j}{\min}(\underline{\bm{\Delta}}_{ij})}
  \end{array}
\right]
\end{align}
The minimum and maximum values for all calculated quantities are stored and then used to rescale the prediction results back to their original scale. Neural network basics are explained in \Cref{apx:NNarchitecture} in more detail.

\subsection{Errors}
Before we look at the numerical results it is crucial to define the errors incurred by the aforementioned approaches. Once the network is trained and the time amplitudes and the shifts are predicted, the errors are as follows:\\
\textbf{POD-NN}:
\begin{equation}\label{eq:err_POD_1}
 E^\mathrm{POD} = \frac{\norm{Q - \tilde{Q}}_{\mathrm{F}}}{\norm{Q}_{\mathrm{F}}} , \quad \text{where} \quad \tilde{Q} = U_r \Sigma_r (V_r)^\top\
\end{equation}
\begin{equation}\label{eq:err_POD_2}
 E^{\mathrm{POD-NN}}_\mathrm{tot} = \frac{\norm{Q - U_r A_\mathrm{NN}}_{\mathrm{F}}}{\norm{Q}_{\mathrm{F}}}
\end{equation}
The error $E^{\mathrm{POD-NN}}_\mathrm{tot}$ is bounded by the POD truncation error $E^\mathrm{POD}$ and the neural network prediction error.
\\
\textbf{sPOD-NN}:
\begin{equation}\label{eq:err_sPOD_2}
    E^\mathrm{sPOD} = \frac{\norm{Q - \tilde{Q}}_{\mathrm{F}}}{\norm{Q}_{\mathrm{F}}} , \quad \text{where} \quad \tilde{Q} \approx \sum^{f}_{k=1} T^{\bm{\Delta}^k}Q^k,  
\end{equation}
\begin{equation}\label{eq:err_sPOD_3}
    E^{\mathrm{sPOD-NN}}_\mathrm{tot} = \frac{\norm{Q - \sum_{k=1}^{f} T^{\bm{\Delta}^k_\mathrm{NN}}\left(U^k_{r^k}  A^k_\mathrm{NN}\right)}_{\mathrm{F}}}{\norm{Q}_{\mathrm{F}}}
\end{equation}
The final error $E^{\mathrm{sPOD-NN}}_\mathrm{tot}$ is bounded by the sPOD truncation error $E^\mathrm{sPOD}$, the neural network prediction error and the error which arises from the shift operations.
\\
\textbf{sPOD-I}:
\begin{equation}\label{eq:err_sPOD_4}
    E^{\mathrm{sPOD-I}}_\mathrm{tot} = \frac{\norm{Q - \sum_{k=1}^{f} T^{\bm{\Delta}^k_\mathrm{I}}\left(U^k_{r^k}  A^k_\mathrm{I}\right)}_{\mathrm{F}}}{\norm{Q}_{\mathrm{F}}}
\end{equation}
The final error $E^{\mathrm{sPOD-I}}_\mathrm{tot}$ is bounded by the sPOD truncation error, the interpolation error, and the error due to shift operations. 

We also look at the error calculated at every time instance \cite{fresca_pod-dl-rom_2022} to gain a comprehensive understanding of the performance of the method. Although this can be computed for all the three methods explained, we only focus on the sPOD-NN method for our analysis. For $(t_j, \mu)$ with \Cref{eq:sPOD_state} we have
\begin{equation}
    E^{\mathrm{sPOD-NN}}_j = \frac{\left\lvert \bm{q}(t_j, \mu) - \bm{\tilde{q}'}(t_j, \mu)\right\rvert}{\sqrt{\frac{1}{N_t}\sum^{N_t}_{j=1}\norm{\bm{q}(t_j, \mu)}^2_{\mathrm{F}}}}\,
\end{equation}
which basically gives us the error at each grid point for a particular $(t_j, \mu)$. 

\section{Numerical results}\label{sec:results}
In order to assess the effectiveness of the sPOD-NN method, we evaluate its numerical performance{\urlstyle{tt} \footnote{The source code is available at \url{https://github.com/MOR-transport/sPOD-NN-paper}}} by the run time and the errors outlined in \Cref{sec:theory}. For a fair comparison, we also compare it with the POD-NN and sPOD-I methods, based on the aforementioned errors. The techniques are evaluated on 1D and 2D wildland fire models. For sPOD we use the algorithm 8 in \cite{krah_non-linear_nodate}. The tuning parameters can be found in \Cref{apx:sPOD_parameters}. 

For benchmarking, the methods were also checked on synthetically generated test data for which the results can be found in \Cref{apx:synthetic_test_case}. The importance of the test case can be inferred from the left plot in \Cref{fig:synthetic_full_err} of \Cref{apx:synthetic_test_case} where we see that $E^{\mathrm{sPOD-I}}_\mathrm{tot}$ decays with the sPOD basis reconstruction error $E^\mathrm{sPOD}$ with an increasing $n_{\mathrm{dof}}$. This implies that if the online phase is substantially accurate then the final error will be dominated only by the basis reconstruction error. This is certainly true for the sPOD-I but not so for the other methods for the given test case. This is by the design of the problem, where the time amplitudes belong to a polynomial space and the method sPOD-I interpolates using polynomials. 

\subsection{Wildland fire model}
We use the model shown in \cite{mandel_wildland_2008}. Considering a domain $\Omega \in \mathbb{R}^d$ with $d = 1, 2$ and a finite time horizon $\mathbb{T}:=[0, t_f]$ where $0 < t_f < \infty$, we compute the temperature $T$ and the fuel supply mass fraction $S$ satisfying the coupled nonlinear PDE
\begin{equation}\label{eq:wildfire}
    \begin{aligned}
    \partial_t T &= \nabla\cdot (k\nabla T) - v\cdot \nabla T + \alpha(Sr(T, \mu, T_a) - \gamma(T - T_a)),\\
    \partial_t S &= -S\gamma_S r(T, \mu, T_a),
    \end{aligned}
\end{equation}
where $\gamma_S r(T, \mu, T_a)$ is called the reaction rate constant which is given by the modified Arrhenius law:
\begin{equation}\label{eq:arrhenius}
    r(T, \mu, T_a) := 
    \begin{cases}
        \mathrm{exp}(-\frac{\mu}{T - T_a}), & \text{if $T > T_a$},\\
        0, & \text{otherwise}\,.
    \end{cases}
\end{equation}
Variables and their fixed values used in the model are shown in \Cref{tab:wildfire_var}.
\begin{table}[h]
\begin{center}
\begin{minipage}{200pt}
\caption{Wildland fire model variables and values for the fixed coefficients involved}\label{tab:wildfire_var}%
\begin{tabular}{@{}llll@{}}
\toprule
Name & Symbol & Unit & Value\\
\midrule
temperature    & $T$   & $\mathrm{K}$  & -  \\
supply mass fraction    & $S$   & -  & -  \\
thermal diffusivity    & $k$   & $\mathrm{m^2 / s}$  & 0.2136  \\
Arrhenius coefficient\footnotemark[1]  & $\mu$   & $\mathrm{K}$  & 558.49  \\
pre-exponential factor    & $\gamma_S$   & $\mathrm{1/s}$  & 0.1625  \\
wind speed    & $v$   & $\mathrm{m/s}$  & 0  \\
temperature rise per second    & $\alpha$   & $\mathrm{K/s}$  & 187.93  \\
scaled heat transfer coefficient    & $\gamma$   & $\mathrm{1/K}$  & $4.8372\times10^{-5}$   \\
ambient temperature    & $T_a$   & $\mathrm{K}$  & 300  \\
\botrule
\end{tabular}
\footnotetext[1]{Tuning factor}
\end{minipage}
\end{center}
\end{table}
For notational convenience, we work with the relative temperature $T - T_a$ instead of the temperature which results in $T_a = 0$ for the rest of the study. 

\subsubsection{1D model}
Here, we will motivate the application of the proposed method to perform non-linear model order reduction of a one-dimensional wildland fire model. As an initial step, we solve the model equations \Cref{eq:wildfire} and \Cref{eq:arrhenius} on a one-dimensional strip of length $l=1000\mathrm{m}$ in a computational domain $\Omega=[0, 1000]$ for $t_f=1400\mathrm{s}$ yielding the time domain $\mathbb{T}=[0, 1400]$. The velocity of wind $v=0$. The time domain is chosen such that the traveling fronts never reach the boundaries. We consider the initial condition to be:
\begin{equation}\label{eq:wildfire1d_init}
    T^0(x):= 1200\mathrm{exp}\left(- \frac{(x-l/2)^2}{200m^2}\right) \quad \text{and} \quad S^0\equiv 1.
\end{equation}
The value of Arrhenius coefficient $\mu=558.49\mathrm{K}$ is used for our simulation. A point to note is that we use $\mu$ as our parameter of choice for tuning the model. A lower $\mu$ results in a faster spread of the fire and a higher $\mu$ results in a slowly expanding fire. Periodic boundary conditions are used for simplicity. The spatial domain is decomposed into $N_x=3000$ grid points making $h_x=1/3 \mathrm{m}$ and discretized with finite difference method with $6^{th}$ order central finite difference stencil. We solve the model for $6000$ time steps. Time integration is done with the standard RK4 method for which $dt$ is calculated with the help of CFL criteria:
\begin{equation}\label{eq:CFL}
    dt = \mathrm{cfl} \cdot \frac{\sqrt{h_x^2 + h_y^2}}{c}
\end{equation}
where we prescribe $\mathrm{cfl} = 0.7$, $h_y=0$ and speed of sound $c=1$.

\begin{figure}[h]
\centering
\includegraphics[scale=0.5]{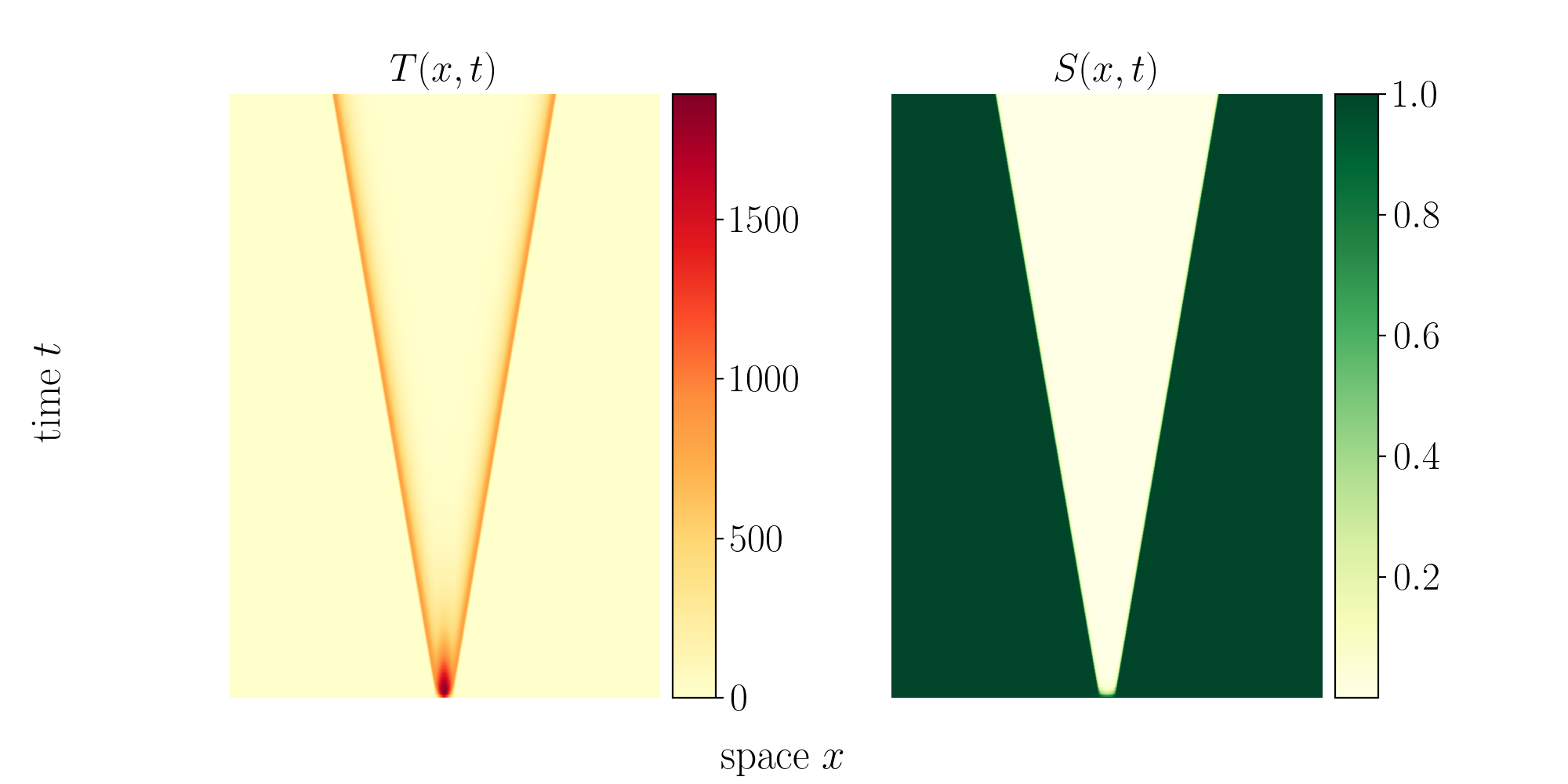}
\caption{Full order model : Temperature(T) and the supply mass fraction(S) profiles computed for $\mu=558.49\mathrm{K}$ }\label{fig:wildfire1d_T_S}
\end{figure}

Subsequently, the obtained temperature and the supply mass fraction profiles are shown in \Cref{fig:wildfire1d_T_S}. In the temperature plot we observe, a localized ignition region from which flame fronts emanate and travel in opposite directions. This is a perfect example of traveling wave-type phenomena. The supply mass fraction varies in the range $[0,1]$ with the region left and right of the two traveling fronts respectively being still unburnt while the central region is completely burnt.

\begin{figure}[h]
\centering
\includegraphics[scale=0.42]{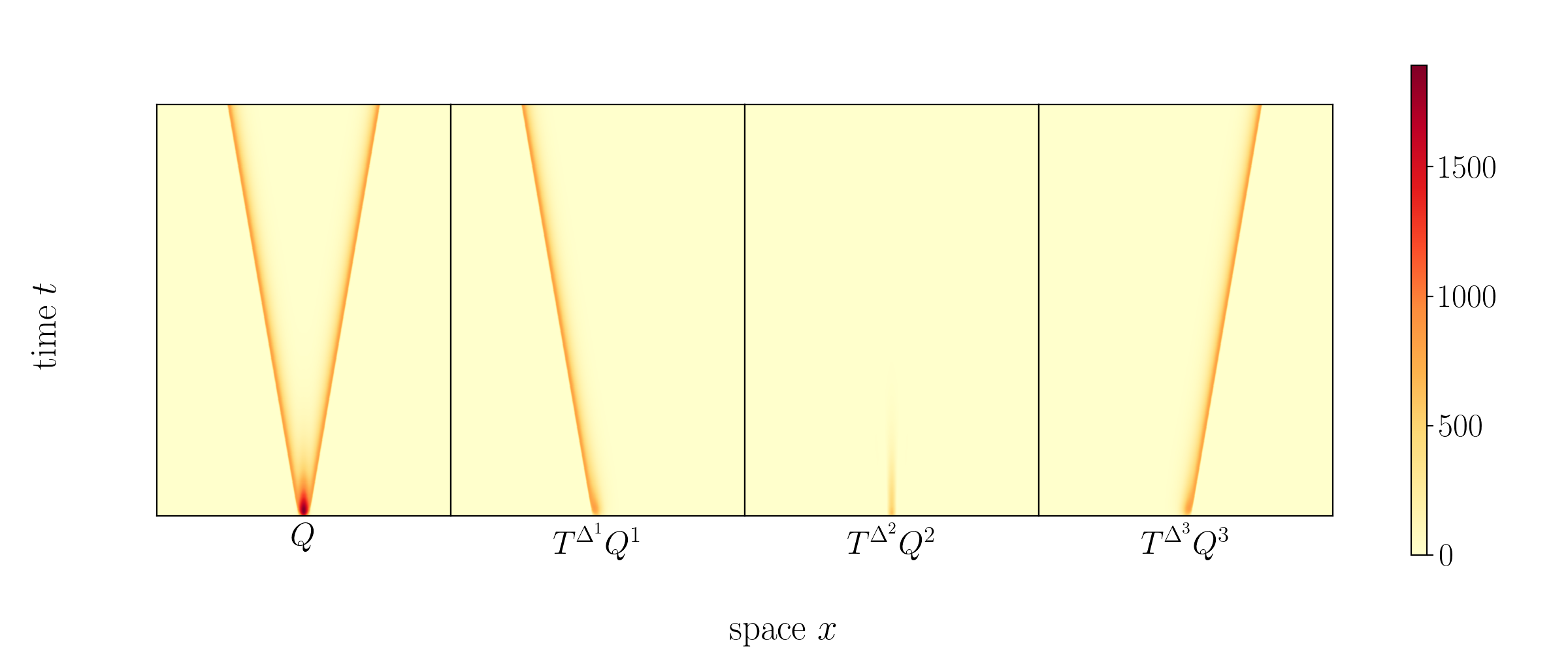}
\caption{sPOD decomposition of the temperature field. The second, third, and fourth plots show the co-moving frames into which the field is decomposed. The transformation operators $T^{\bm{\Delta}^k}$ transform the stationary decomposed frames into the co-moving frames which when combined gives the reconstructed snapshot shown in the first plot $$Q = T^{\bm{\Delta}^1} Q^1 + T^{\bm{\Delta}^2} Q^2 + T^{\bm{\Delta}^3} Q^3$$}\label{fig:wildfire1d_sPOD_decomp}
\end{figure}

For constructing non-linear parametric ROMs we solve \Cref{eq:wildfire} and \Cref{eq:arrhenius} for $\mu\in\{540\mathrm{K}, 550\mathrm{K}, 560\mathrm{K}, 570\mathrm{K}, 580\mathrm{K}\}$ and sample naively every $10^{th}$ snapshot in time to construct a parametric snapshot matrix $Q$. More advanced sampling strategies can be found in \cite{karcher_adaptive_2022}. With $N_t=6000/10=600$ time instances, $N_p=5$ parameter instances of $\mu$ and $M=N_x=3000$ the constructed matrix $Q\in \mathbb{R}^{M\times N_t N_p} = \mathbb{R}^{3000\times 3000}$. Once the parameter snapshot matrix is assembled we perform dimensionality reduction by sPOD and obtain a low-dimensional representation of the data as shown in \Cref{fig:wildfire1d_sPOD_decomp}. The number of co-moving frames is decided by the user and for this example it is chosen as $3$ (one for each of the left and right traveling fronts and one for the initial ignition regime). The shifts for the frames required for sPOD are calculated by tracking the gradients of the field variable as explained in \cite{black_efficient_2021}. We note here that we only show the results for the temperature field for simplicity, however, the entire procedure can be replicated for the supply mass fraction as well, the results of which could be found in \Cref{apx:results_supply_mass_fraction}. The motivation for using sPOD is substantiated by the basis reconstruction error results shown in \Cref{tab:1dwildfire_errors}. We observe that as the number of modes increases the drop in $E^{\mathrm{sPOD}}$ is significant in comparison to $E^{\mathrm{POD}}$. For eg. just for $16+10+16=42$ modes we could get to $E^{\mathrm{sPOD}}\sim \mathcal{O}(10^{-4})$ whereas $E^{\mathrm{POD}}\sim \mathcal{O}(10^{-2})$. 
% The shift transformation error $\sim \mathcal{O}(10^{-7})$ thus is neglected hereafter. 
\begin{table}[h]
\begin{center}
\begin{minipage}{\textwidth}
\caption{Offline and online error study w.r.t number of modes for 1D wildland fire model}\label{tab:1dwildfire_errors}
\begin{tabular*}{\textwidth}{@{\extracolsep{\fill}}lcccccc@{\extracolsep{\fill}}}
\toprule%
Modes\footnotemark[1]& \multicolumn{2}{@{}c@{}}{Offline errors} & \multicolumn{3}{@{}c@{}}{Online errors} \\\cmidrule{2-3}\cmidrule{4-6}%
$r_1+r_2+r_3$ & $E^{\mathrm{sPOD}}$ & $E^{\mathrm{POD}}$ & $E^{\mathrm{sPOD-NN}}_{\mathrm{tot}}$ & $E^{\mathrm{sPOD-I}}_{\mathrm{tot}}$ & $E^{\mathrm{POD-NN}}_{\mathrm{tot}}$ \\
\midrule
1 + 1 + 1 & $3.85\times 10^{-1}$ & $6.90\times 10^{-1}$ & $3.93\times 10^{-1}$ & $2.36\times 10^{-1}$ & $6.92\times 10^{-1}$ \\
2 + 2 + 2 & $1.97\times 10^{-1}$ & $5.03\times 10^{-1}$ & $2.34\times 10^{-1}$ & $2.31\times 10^{-1}$ & $5.03\times 10^{-1}$ \\
4 + 2 + 4 & $6.13\times 10^{-2}$ & $3.57\times 10^{-1}$ & $1.53\times 10^{-1}$ & $7.83\times 10^{-2}$ & $3.55\times 10^{-1}$ \\
6 + 2 + 6 & $2.49\times 10^{-2}$ & $2.69\times 10^{-1}$ & $1.00\times 10^{-1}$ & $4.05\times 10^{-2}$ & $2.67\times 10^{-1}$ \\
8 + 4 + 8 & $1.26\times 10^{-2}$ & $1.86\times 10^{-1}$ & $7.25\times 10^{-2}$ & $3.83\times 10^{-2}$ & $1.85\times 10^{-1}$ \\
10 + 4 + 10 & $5.51\times 10^{-3}$ & $1.48\times 10^{-1}$ & $4.09\times 10^{-2}$ & $3.67\times 10^{-2}$ & $1.48\times 10^{-1}$ \\
14 + 8 + 14 & $1.20\times 10^{-3}$ & $8.07\times 10^{-2}$ & $3.99\times 10^{-2}$ & $3.64\times 10^{-2}$ & $8.67\times 10^{-2}$ \\
16 + 10 + 16 & $6.94\times 10^{-4}$ & $6.06\times 10^{-2}$ & $3.78\times 10^{-2}$ & $3.64\times 10^{-2}$ & $7.85\times 10^{-2}$ \\
\botrule
\end{tabular*}
\footnotetext[1]{We also have $r=r_1 + r_2 + r_3 + 2$ where $r_1$ and $r_3$ are the ranks of the co-moving frames and $r_2$ is the rank of the stationary frame and $2$ accounts for two additional degrees of freedom controlling the shifts. For the comparison, we, therefore, use $r$ modes for POD.}
\end{minipage}
\end{center}
\end{table}

\begin{figure}[h]
\centering
\includegraphics[scale=0.38]{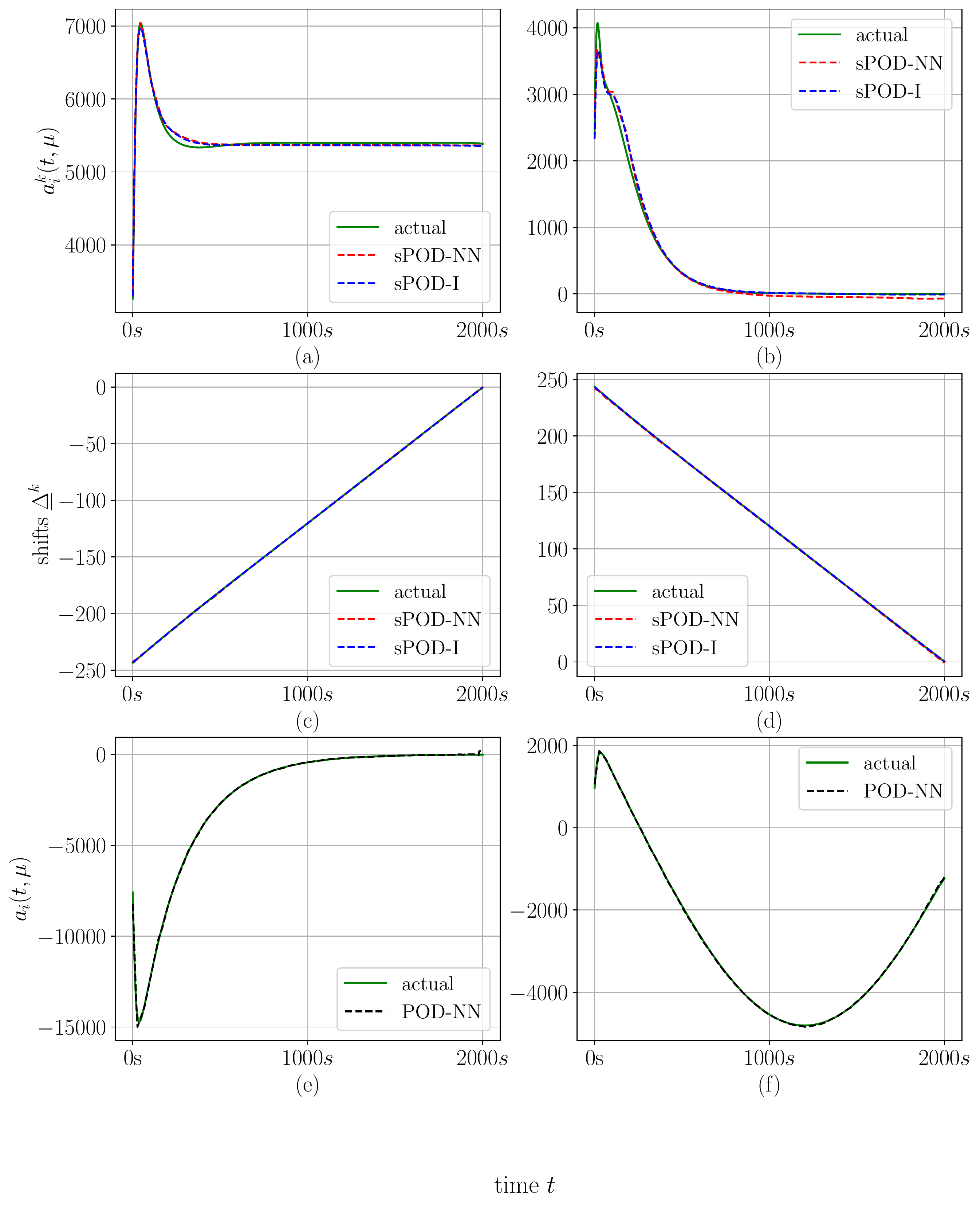}
\caption{Plot (a) and (b) show the first modes for the frame 1 and 2 respectively as frame 3 is identical to its counterpart frame 1. (c) and (d) show the shifts for frame 1 and frame 3 respectively. (e) and (f) show the first two modes corresponding to the POD-NN approach.}\label{fig:wildfire1d_time_amplitudes_shifts_pred}
\end{figure}

\begin{figure}[h]
\centering
\includegraphics[scale=0.40]{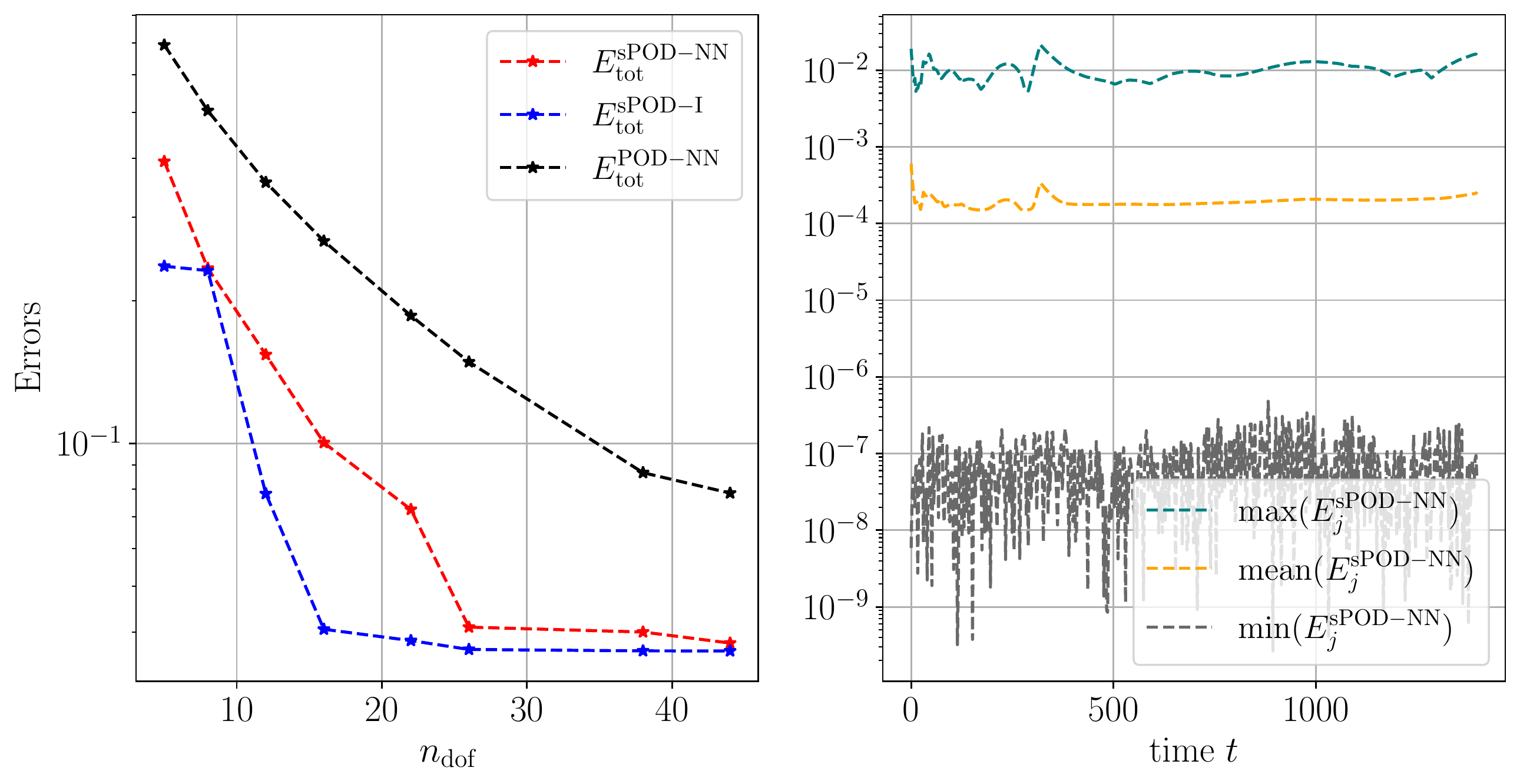}
\caption{The first plot shows the error decay for an increasing $n_{\mathrm{dof}}$. The errors labeled $E_{\mathrm{tot}}$ show the final reconstruction error after prediction or interpolation. The second plot shows the trend of the relative error $E^{\mathrm{sPOD-NN}}_j$ over time.}\label{fig:wildfire1d_full_err}
\end{figure}

As a result of the dimensionality reduction we obtain the time amplitude matrix $A^k$ from $Q^k$ as shown in \Cref{eq:sPOD_extract} and we already have the shifts calculated for all $k$. Following the procedure outlined in \Cref{sec:theory} for training the neural network model we assemble the time amplitude matrix $\hat{A}\in \mathbb{R}^{42 \times 3000}$ and the parameter matrix $P\in \mathbb{R}^{2 \times 3000}$. We only consider $2$ shifts in the aforementioned assembly of the matrix $\hat{A}$, as the shift for the second frame $\underline{\bm{\Delta}}^2 = 0$. The network is trained for $N_{\mathrm{epochs}}=100000$ with batch size $N_b=100$. Early stopping criteria are imposed to prevent overfitting which stops the training if the validation loss does not decrease for 3000 consecutive epochs. For testing, we choose $\mu = [558.49\mathrm{K}]$ and use the trained model to predict the time amplitudes and the shifts which are shown in \Cref{fig:wildfire1d_time_amplitudes_shifts_pred}. For a more quantitative picture, we look at the error estimates shown in \Cref{fig:wildfire1d_full_err}.
In the first plot, we see that as the $n_{\mathrm{dof}}$ increases all three errors go down as expected. However, for both sPOD-NN and sPOD-I the $E_{\mathrm{tot}}$ starts to stagnate after a certain $n_{\mathrm{dof}}$. We observe that when we increase the $n_{\mathrm{dof}}$ the amount of new information added to the training data is not substantial. On the other hand, the parameter matrix $P$ remains the same. Thus in turn the approximation capability of the network remains almost the same even while making the prediction task more and more difficult. This also seems to be the case for sPOD-I. For POD-NN this effect can not be fully captured here although, slight stagnation could be seen towards the end of the curve. This effect is also mentioned in Fresca et.al (2021) \cite{fresca_pod-dl-rom_2022}. sPOD-NN is able to reach to an $E^{\mathrm{sPOD-NN}}_{\mathrm{tot}} \sim 0.037$. We also study the error $E^{\mathrm{sPOD-NN}}_j$ which is shown in the right plot of \Cref{fig:wildfire1d_full_err} where we see $\max  \sim \mathcal{O}(10^{-2})$ whereas mean $ < \mathcal{O}(10^{-3})$. 

\begin{figure}[h]
\centering
\includegraphics[scale=0.38]{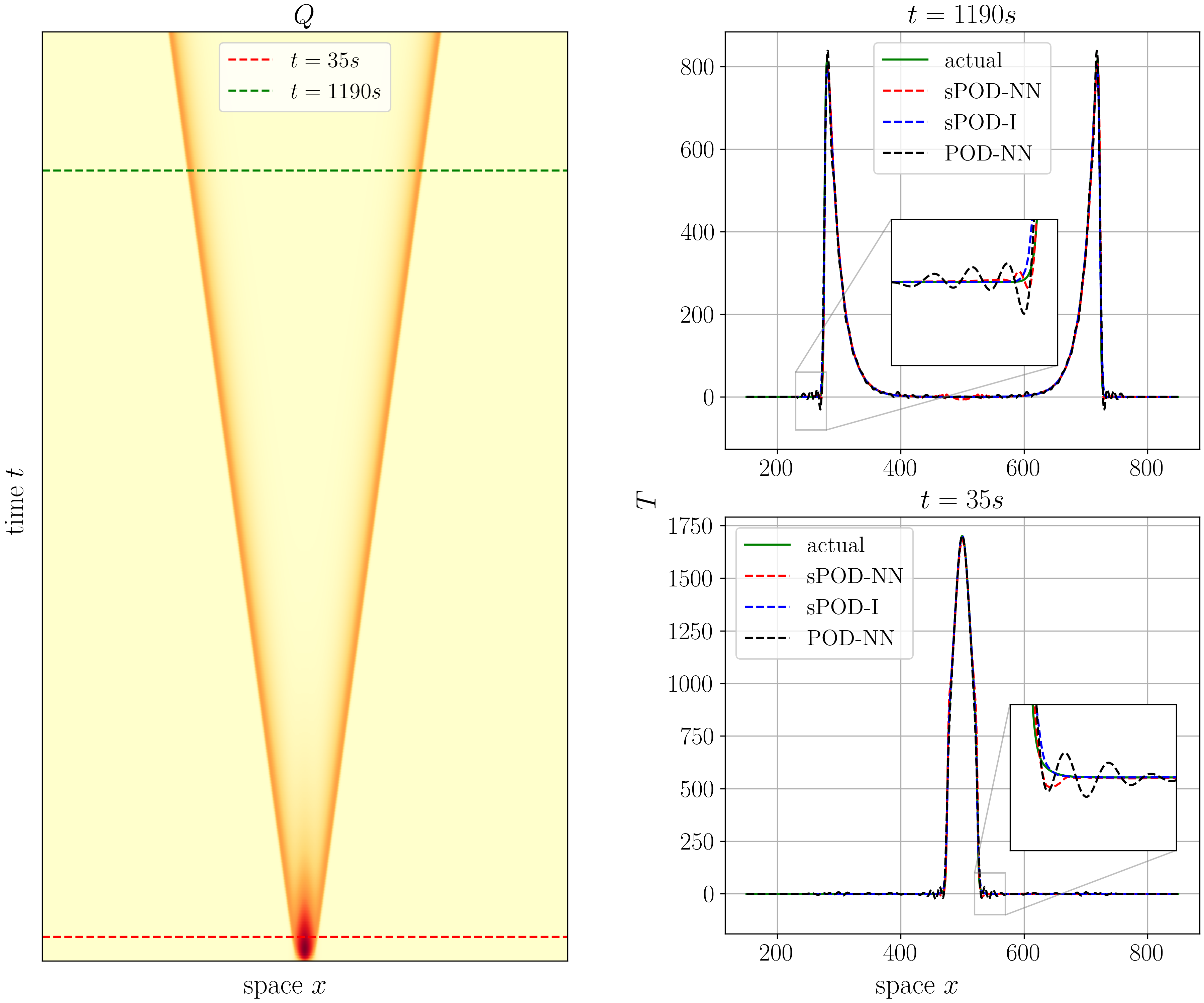}
\caption{This figure captures the prediction accuracy of the studied methods. We study the cross-sectional views of the temperature profile at $t=35s$ and $1190s$.}\label{fig:wildfire1d_T_x_cs}
\end{figure}

The full reconstructed results are plotted in \Cref{fig:wildfire1d_T_x_cs}. We select two random time instances: one near the ignition and the other nearing the end of the time domain and the prediction results are shown for all the methods at these instances. We could observe that the POD-NN although being able to have $E^{\mathrm{POD-NN}}_{\mathrm{tot}} \sim 0.078$, has spurious oscillations. The $E^{\mathrm{POD}} \sim 0.06$ thus we infer that the prediction accuracy for POD-NN is limited by the basis reconstruction error itself, which cannot be resolved unless more modes are added for the study. As for sPOD-NN and sPOD-I, we see minor oscillations near the sharp edges of the curve.  

\subsubsection{2D model (\textit{without wind})} 
\begin{figure}[h]
\centering
\includegraphics[scale=0.5]{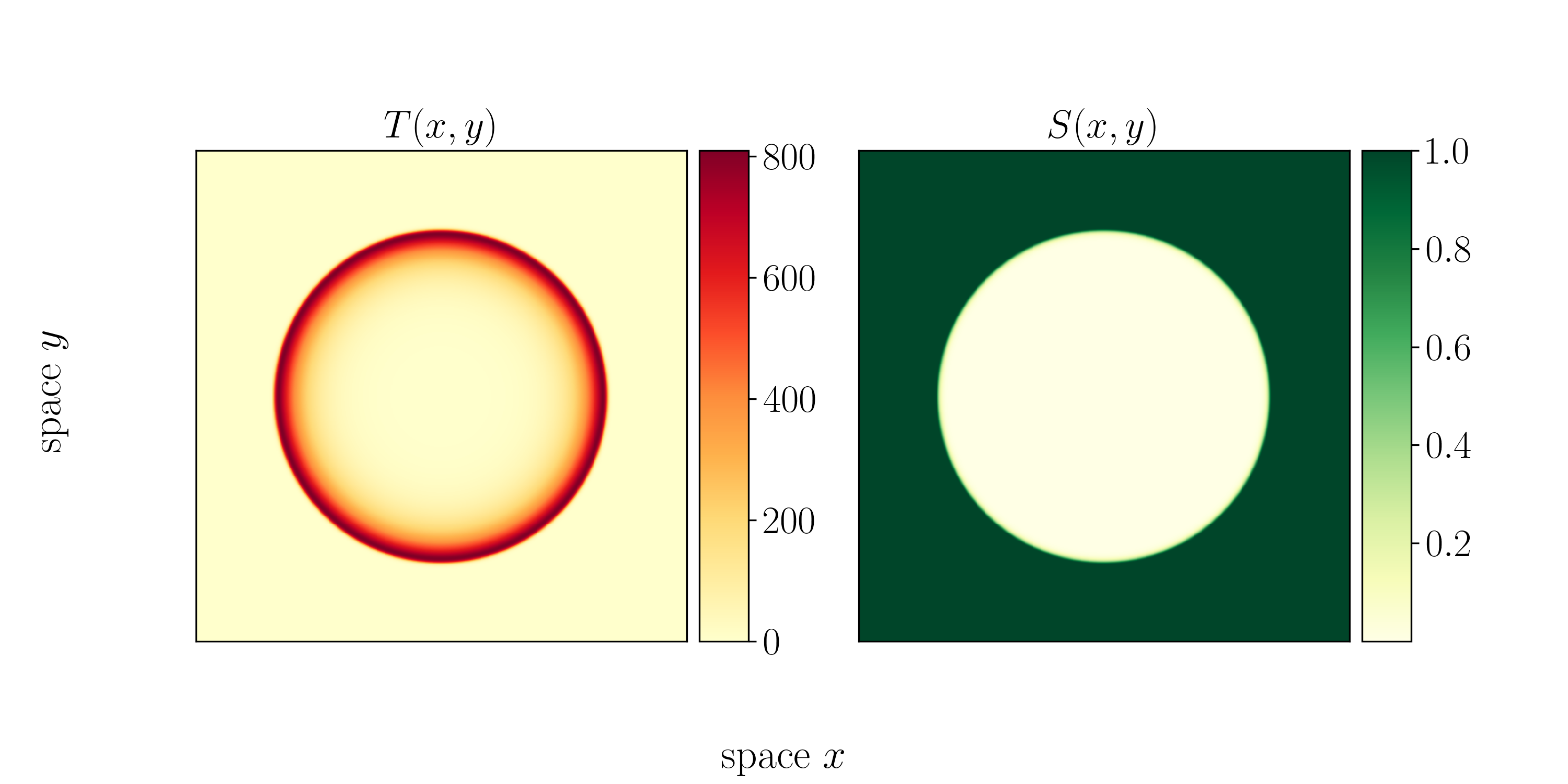}
\caption{Full order Model: Temperature(T) and supply mass fraction(S) profile at $t=1000s$.}\label{fig:wildfire2d_T_S}
\end{figure}

The model equations for the wildland fire can also be extended to a 2D case. We solve the equations \Cref{eq:wildfire} and \Cref{eq:arrhenius} on a two-dimensional square of side length $l=500$ in a computational domain $\Omega = [0, 500] \times [0, 500]$ for $t_f = 1000s$, yielding the time domain $\mathbb{T} = [0, 1000]$ such that the traveling fronts do not reach the boundaries. The velocity of wind $v=(0, 0)$ results in no change in the topology of the fronts. The initial condition is given to be:
\begin{equation}\label{eq:wildfire2d_init}
    T^0(x, y):= 1200\mathrm{exp}\left(- \left(\frac{(x-l/2)^2}{200m^2} + \frac{(y-l/2)^2}{200m^2}\right)\right) \quad \text{and} \quad S^0(x, y)\equiv 1.
\end{equation}
Periodic boundary conditions are employed. The spatial domain is split into $N_x=500, N_y=500$ grid points making $h_x=1, h_y=1$ and discretized with finite difference method with $6^{th}$ order central finite difference stencil. We solve the system for $1000$ time steps. Time integration is done with standard RK4 scheme and the $dt$ is computed by CFL condition shown in \Cref{eq:CFL} where we prescribe $\mathrm{cfl} = 0.7$ and $c=1$. Similar to the 1D case we consider $\mu=558.49\mathrm{K}$ which also acts as a tuning parameter in the model reduction procedure. The temperature and the supply mass fraction profiles are shown in \Cref{fig:wildfire2d_T_S}.

For constructing non-linear parametric ROMs we follow the same steps as mentioned in the 1D case, solving for $\mu\in\{540\mathrm{K}, 550\mathrm{K}, 560\mathrm{K}, 570\mathrm{K}, 580\mathrm{K}\}$ and sampling every $10^{th}$ snapshot in time to construct a parametric snapshot matrix $Q$. With $N_t=1000/10=100$ time instances, $N_p=5$ parameter instances of $\mu$ and $M=N_x * N_y=250000$ the constructed matrix $Q\in \mathbb{R}^{M\times N_t N_p} = \mathbb{R}^{250000\times 500}$. We then carry out the dimensionality reduction through sPOD. We, however, need the shifts prior to running the sPOD algorithm. To this end, we observe that the 2D problem is set up in a way that the developed flame front is perfectly circular throughout the domain. Because of this radial symmetry, we can transform the data from a Cartesian to a polar coordinate system which simplifies the problem significantly. For performing model reduction in polar coordinate system we have $\underline{\bm{\Delta}}^k = (\underline{\Delta}^k_{\mathrm{R}}, \underline{\Delta}^k_{\theta})$ as shifts.

\begin{figure}[h]
\centering
\includegraphics[scale=0.4]{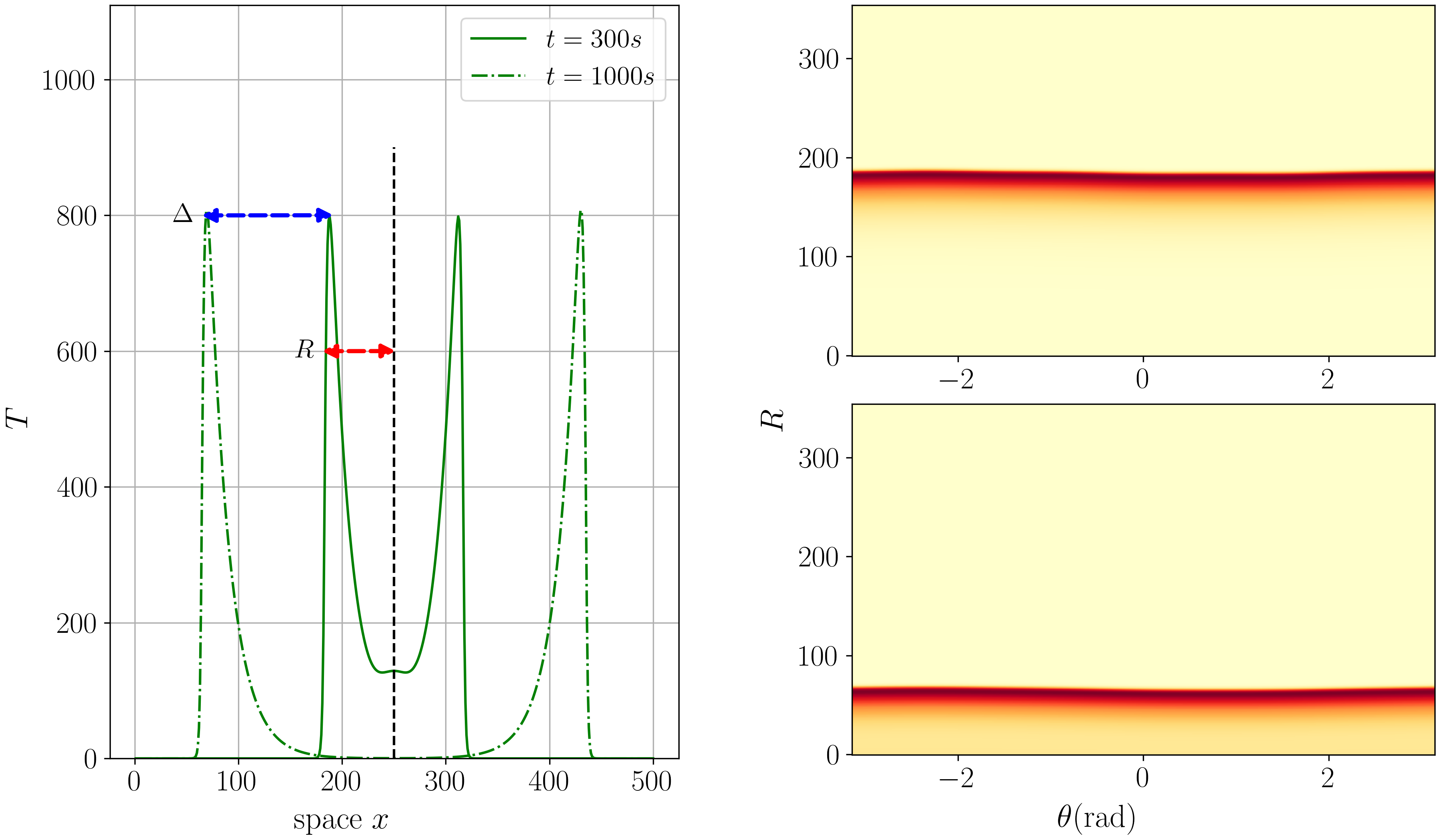}
\caption{The left plot shows the cross-sectional view of the temperature profile for two different time instances. The other two plots on the right show the temperature profiles at the mentioned time instances in the polar coordinate system.}\label{fig:wildfire2d_polar_cs}
\end{figure}

\begin{figure}[h]
\centering
\includegraphics[scale=0.4]{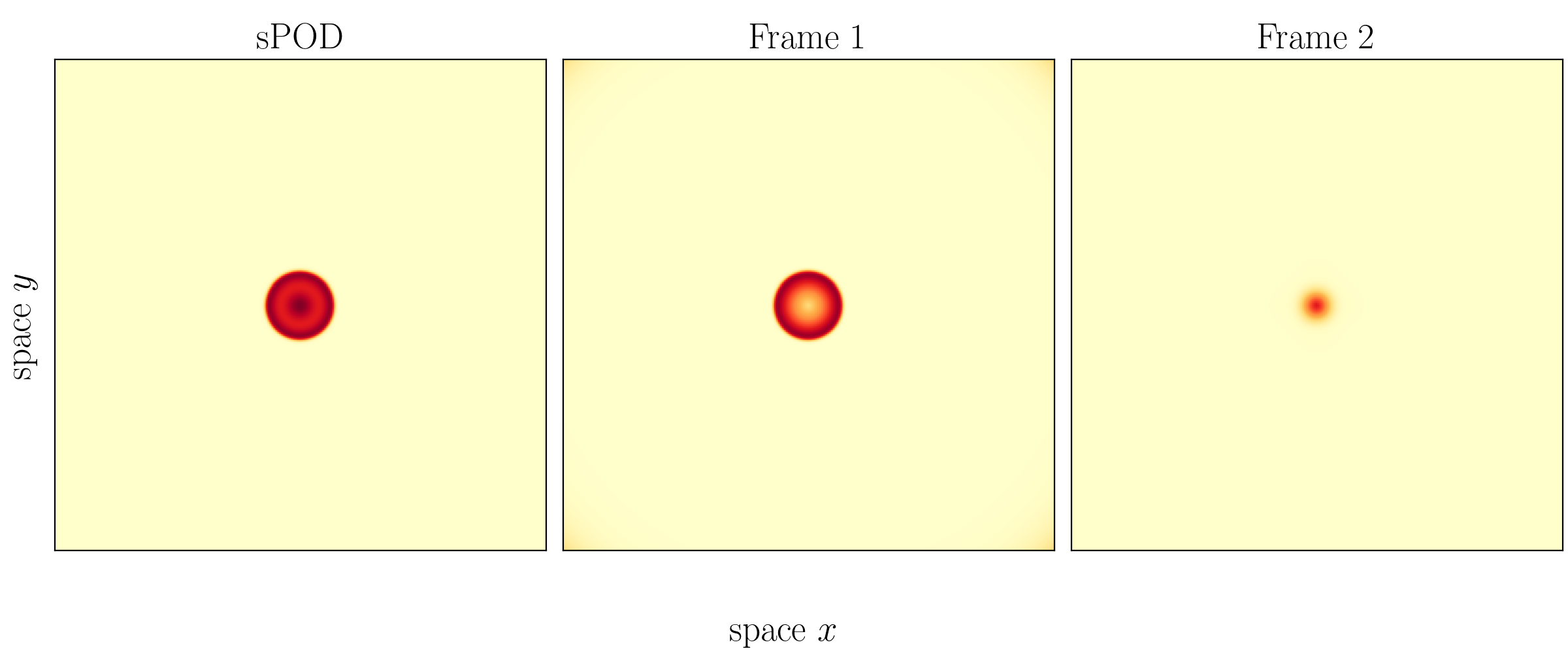}
\caption{This figure shows the sPOD decomposition of the temperature field at $t=100s$. The first plot shows the sPOD reconstruction whereas we observe the decomposition of the field into frame 1 consisting of the traveling wave part and frame 2 containing the initial ignition. }\label{fig:wildfire2d_sPOD_frames}
\end{figure}

Let us consider the \Cref{fig:wildfire2d_polar_cs}, the first plot shows the 1D cross-sectional views of the temperature profile at two different time instances. $t=1000s$ is the final instance that can be considered as the reference position. The $\Delta$ is given as:
\begin{equation*}
    \Delta = \underbrace{x(t=300s)}_{\mathrm{R}} - x(t=1000s)
\end{equation*}
There is no dependency of the shifts on the $\theta$ coordinate. The $\Delta$ only depends upon $\mathrm{R}$ and $t$. The other two plots show the temperature profiles in the polar coordinate system for the two-time instances $t=300s$ and $t=1000s$. 
With sPOD, we now aim at decomposing the polar temperature field into two frames: the first one captures the traveling fronts, and the second one captures the initial ignition regime. For this we consider, $\underline{\Delta}^1_{\mathrm{R}} = \Delta, \: \underline{\Delta}^1_{\theta} = 0$ and $\underline{\bm{\Delta}}^2 = (0, 0)$. The sPOD decomposition of the temperature is shown in \Cref{fig:wildfire2d_sPOD_frames} in Cartesian coordinates. 
\begin{table}[h]
\begin{center}
\begin{minipage}{\textwidth}
\caption{Offline and online error study w.r.t number of modes for 2D wildland fire model(without wind)}\label{tab:2dwildfire_errors}
\begin{tabular*}{\textwidth}{@{\extracolsep{\fill}}lcccccc@{\extracolsep{\fill}}}
\toprule%
Modes\footnotemark[1]& \multicolumn{2}{@{}c@{}}{Offline errors} & \multicolumn{3}{@{}c@{}}{Online errors} \\\cmidrule{2-3}\cmidrule{4-6}%
$r_1+r_2$ & $E^{\mathrm{sPOD}}$ & $E^{\mathrm{POD}}$ & $E^{\mathrm{sPOD-NN}}_{\mathrm{tot}}$ & $E^{\mathrm{sPOD-I}}_{\mathrm{tot}}$ & $E^{\mathrm{POD-NN}}_{\mathrm{tot}}$ \\
\midrule
1 + 1 & $3.67\times 10^{-1}$ & $6.20\times 10^{-1}$ & $2.02\times 10^{-1}$ & $2.05\times 10^{-1}$ & $6.17\times 10^{-1}$ \\
5 + 2 & $2.18\times 10^{-2}$ & $3.51\times 10^{-1}$ & $3.20\times 10^{-2}$ & $3.23\times 10^{-2}$ & $3.46\times 10^{-1}$ \\
7 + 2 & $1.13\times 10^{-2}$ & $2.89\times 10^{-1}$ & $3.22\times 10^{-2}$ & $2.99\times 10^{-2}$ & $2.84\times 10^{-1}$ \\
10 + 3 & $5.25\times 10^{-3}$ & $2.03\times 10^{-1}$ & $3.79\times 10^{-2}$ & $2.89\times 10^{-2}$ & $2.01\times 10^{-1}$ \\
16 + 4 & $1.88\times 10^{-3}$ & $1.18\times 10^{-1}$ & $2.35\times 10^{-2}$ & $2.85\times 10^{-2}$ & $1.25\times 10^{-1}$ \\
19 + 5 & $1.33\times 10^{-3}$ & $8.91\times 10^{-2}$ & $1.77\times 10^{-2}$ & $2.85\times 10^{-2}$ & $9.85\times 10^{-2}$ \\
\botrule
\end{tabular*}
\footnotetext[1]{We also have $r=r_1 + r_2 + 1$, where $r_1$ is the rank of the moving frame, $r_2$ is the rank of the stationary frame, and $1$ accounts for one additional degree of freedom controlling the shift. For comparison, we therefore use $r$ modes for the POD.}
\end{minipage}
\end{center}
\end{table}

The basis reconstruction error results are shown in \Cref{tab:2dwildfire_errors}. We observe that as we keep on adding more modes for the reconstruction the drop in $E^{\mathrm{sPOD}}$ is significant in comparison to $E^{\mathrm{POD}}$. For eg. sPOD converges for $19+5=24$ modes to $E^{\mathrm{sPOD}}\sim \mathcal{O}(10^{-3})$ whereas $E^{\mathrm{POD}}\sim \mathcal{O}(10^{-2})$.

\begin{figure}[h]
\centering
\includegraphics[scale=0.38]{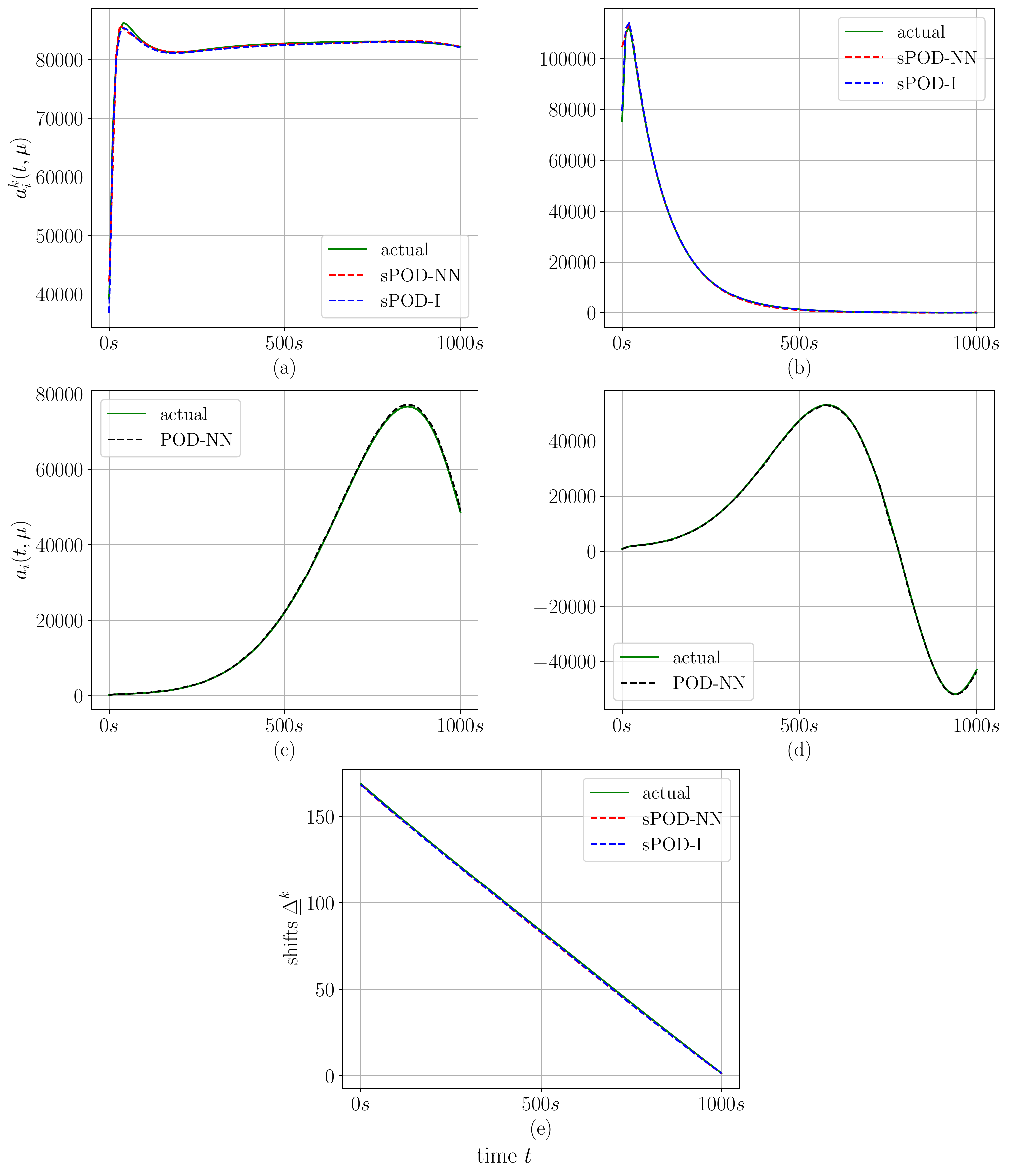}
\caption{Plot (a) and (b) show the time amplitude predictions of the first mode for frame 1 and frame 2 respectively. (c) and (d) show the prediction results for the first two modes corresponding to the POD-NN approach. (e) show the shift prediction corresponding to the sPOD-NN approach.}\label{fig:wildfire2d_time_amplitudes_shifts_pred}
\end{figure}

\begin{figure}[h]
\centering
\includegraphics[scale=0.40]{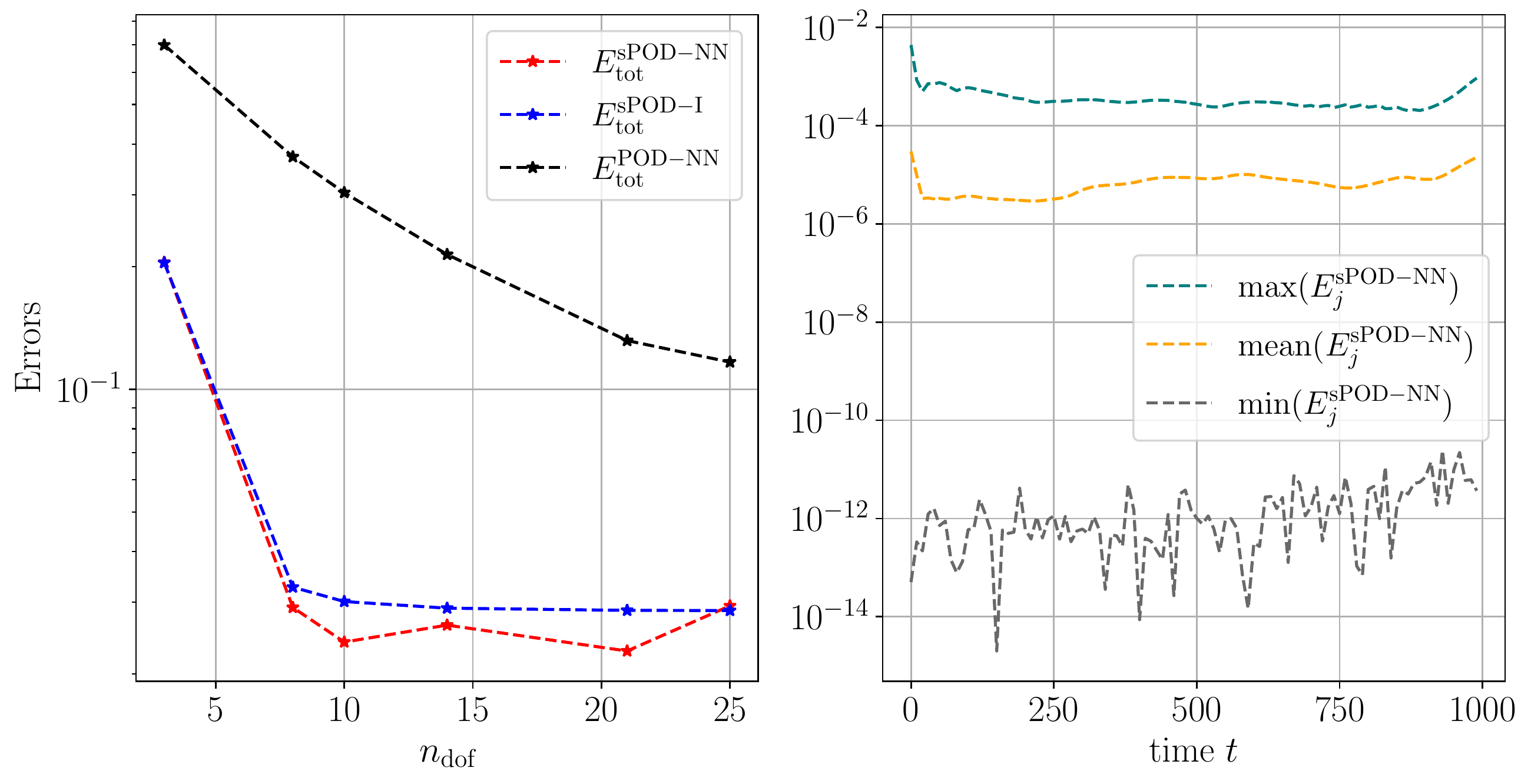}
\caption{The first plot shows the error decay for an increasing $n_{\mathrm{dof}}$. The errors labeled $E_{\mathrm{tot}}$ show the final reconstruction error after prediction or interpolation. The second plot shows the trend of the relative error $E^{\mathrm{sPOD-NN}}_j$ over time.}\label{fig:wildfire2d_full_err}
\end{figure}

\begin{figure}[h]
\centering
\includegraphics[scale=0.40]{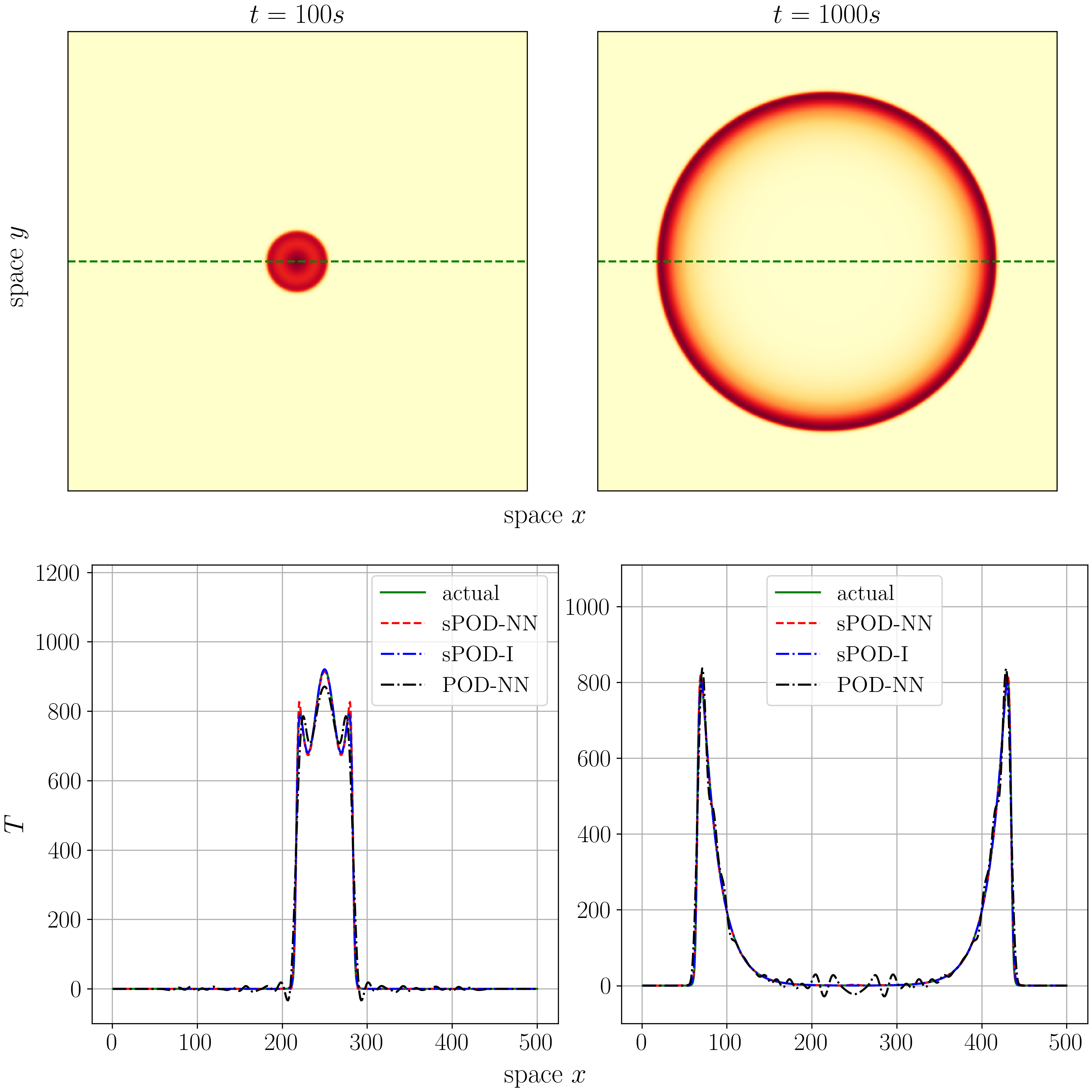}
\caption{The top two plots show the temperature profiles at $t=100s$ and $t=1000s$. The bottom two plots show the cross-sectional view of their top counterparts with the prediction results for sPOD-NN and POD-NN procedure.}\label{fig:wildfire2d_T_x_cs}
\end{figure}

As a result of the sPOD, we obtain the time amplitudes for both the frames and we subsequently construct the time amplitude matrix $\hat{A}$ and the parameter matrix $P$ for training as shown in \Cref{sec:theory}. For this problem $\hat{A} \in \mathbb{R}^{25 \times 500}$ and $P \in \mathbb{R}^{2 \times 500}$. The network is trained for $N_{\mathrm{epochs}}=200000$ with batch size $N_b=50$. Early stopping criteria are imposed to prevent overfitting, which stops the training if the validation loss does not decrease for 4000 consecutive epochs. For testing, we choose $\mu = [558.49\mathrm{K}]$ to predict the time amplitudes and the shifts which are shown in \Cref{fig:wildfire2d_time_amplitudes_shifts_pred}. Quantitative error estimates are shown in \Cref{fig:wildfire2d_full_err} where in the first plot we observe the similar stagnation behavior that was observed in the 1D wildland fire model case. The $E^{\mathrm{sPOD-NN}}_{\mathrm{tot}} \sim 0.017$. In the second plot, we see the error over the whole time interval $E^{\mathrm{sPOD-NN}}_j$ where we see $\max  < \mathcal{O}(10^{-2})$ and the mean $ < \mathcal{O}(10^{-4})$.
Full reconstruction results are shown in \Cref{fig:wildfire2d_T_x_cs}. Here we see the cross-sectional view of the temperature profile at two different time instances. We observe that the POD-NN has spurious oscillations. The $E^{\mathrm{POD}} \sim 0.089$ and $E^{\mathrm{POD-NN}}_{\mathrm{tot}} \sim 0.098$ thus we infer that the prediction accuracy for POD-NN is once again limited by the basis reconstruction error itself. As for sPOD-NN and sPOD-I, we see very minor oscillations near the sharp edges of the curve.  

\subsubsection{2D model (\textit{with wind})}
\label{sec:2dwithwind}
\begin{figure}[h]
\centering
\includegraphics[scale=0.5]{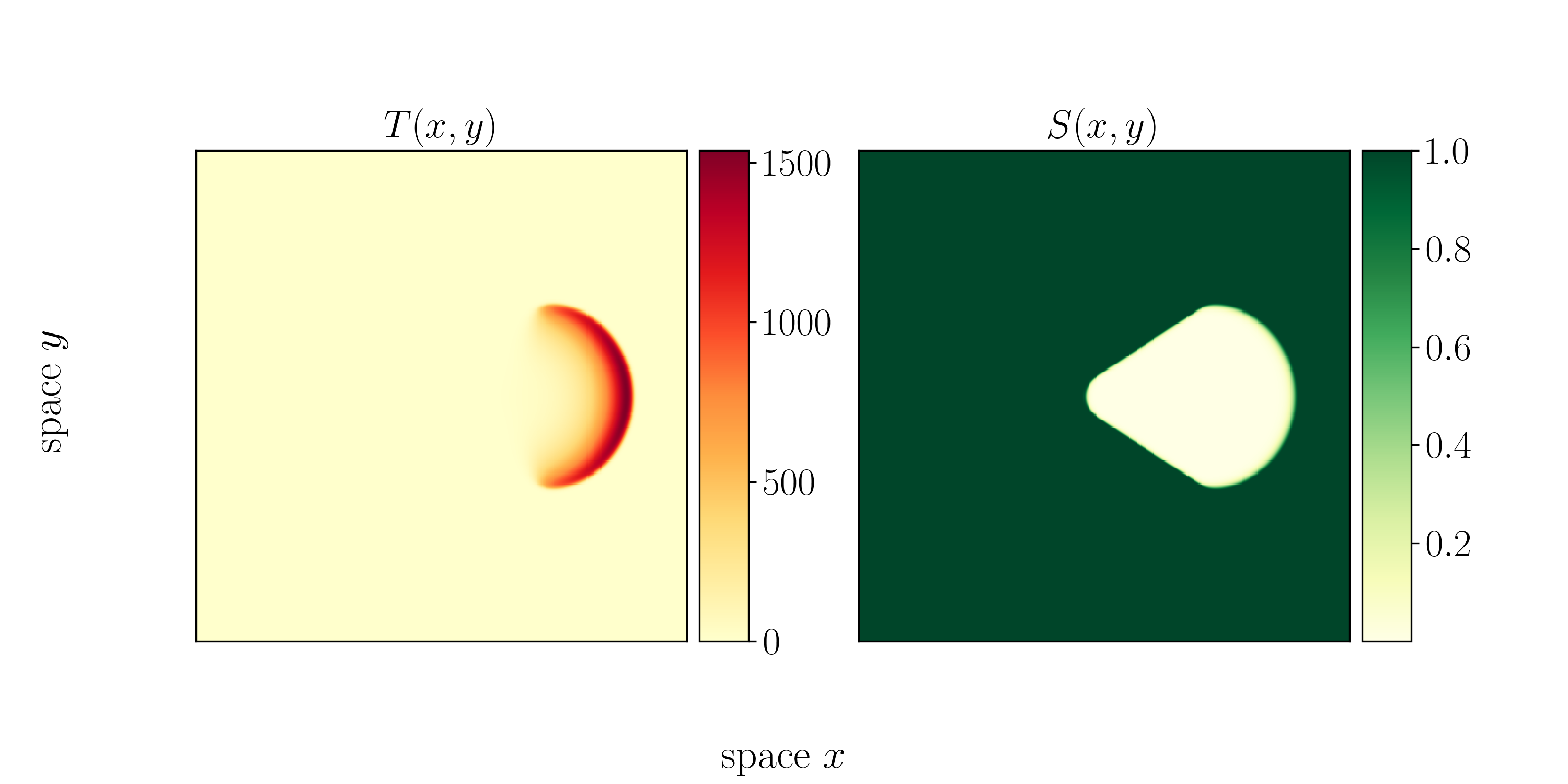}
\caption{Full order model: Temperature(T) and supply mass fraction(S) profile at $t=500s$.}\label{fig:wildfire2dnonlinear_T_S}
\end{figure}

We consider the same model equations as in the previous case albeit with a minor modification in \Cref{eq:wildfire} where we prescribe $v=(0.2 \:\mathrm{m/s}, 0)$ which is the wind velocity applied only in the $x$ direction. The two-dimensional square domain is kept the same as the previous case and the spatial domain is split into $N_x=500, N_y=500$ grid points making $h_x=1, h_y=1$ and discretized with finite difference method with $6^{th}$ order central finite difference stencil. We solve the system for $500$ time steps for $t_f = 500 s$. The temperature and the supply mass fraction profiles are shown in \Cref{fig:wildfire2dnonlinear_T_S}.

\begin{figure}[h]
\centering
\includegraphics[scale=0.4]{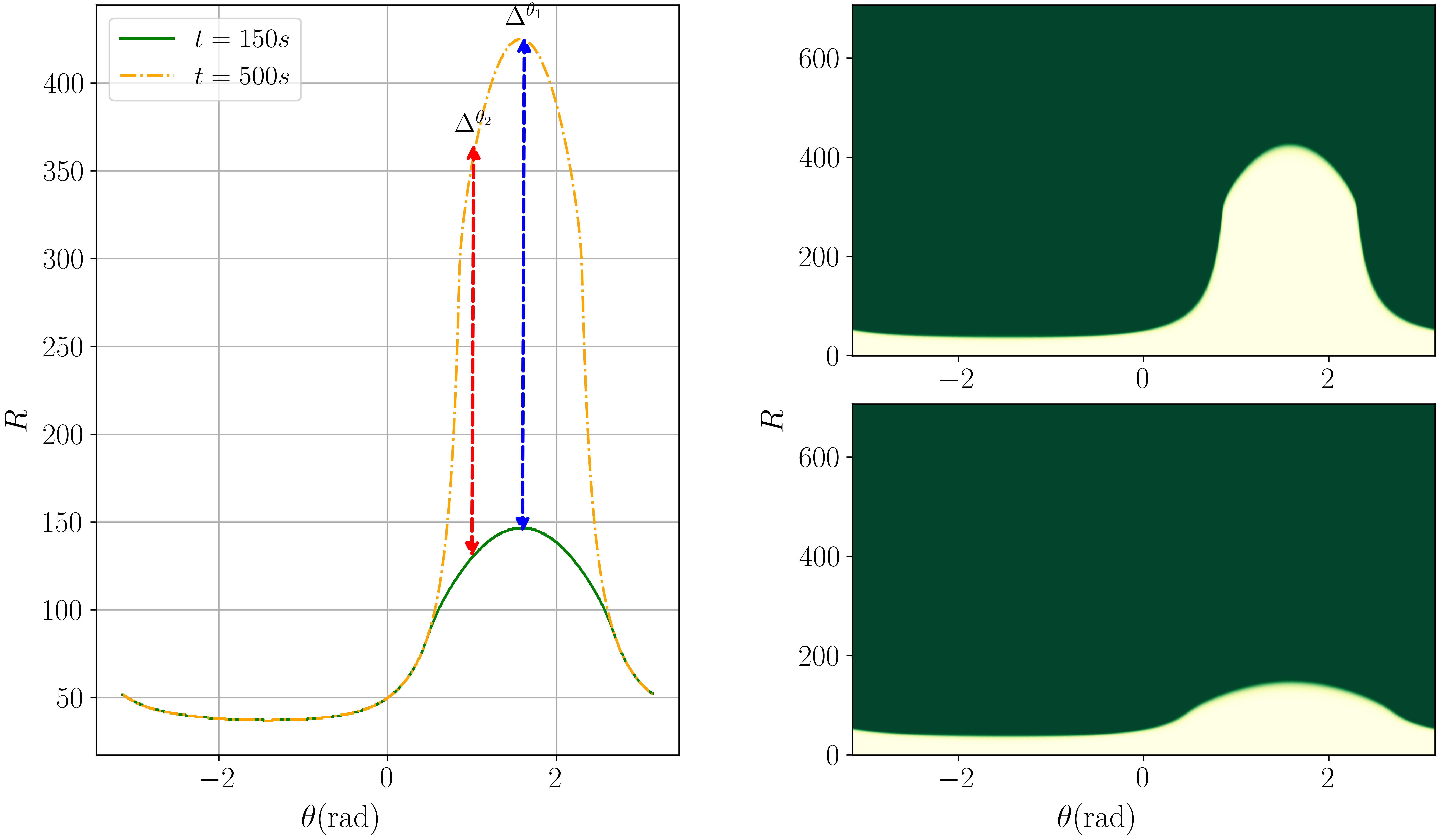}
\caption{The plots show the supply mass fractions at two different time instances in the polar coordinate system.}\label{fig:wildfire2d_polar_nonlinear_cs}
\end{figure}

\begin{figure}[h]
\centering
\includegraphics[scale=0.4]{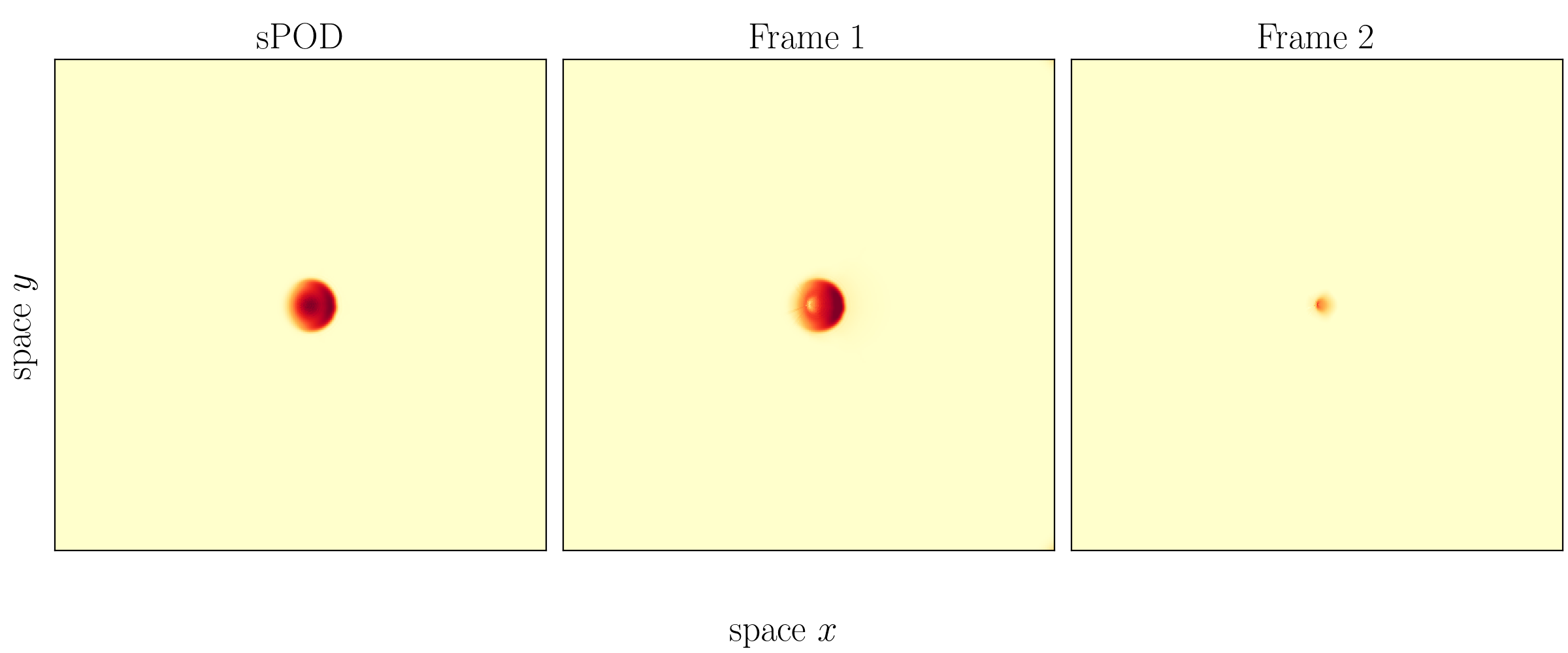}
\caption{This figure shows the sPOD decomposition of the temperature field at $t=50s$. The first plot shows the sPOD reconstruction and the respective decomposition into the traveling part and the stationary initial ignition part are shown in the second and the third plot.}\label{fig:wildfire2dnonlinear_sPOD_frames}
\end{figure}

\begin{figure}[h]
\centering
\includegraphics[scale=0.43]{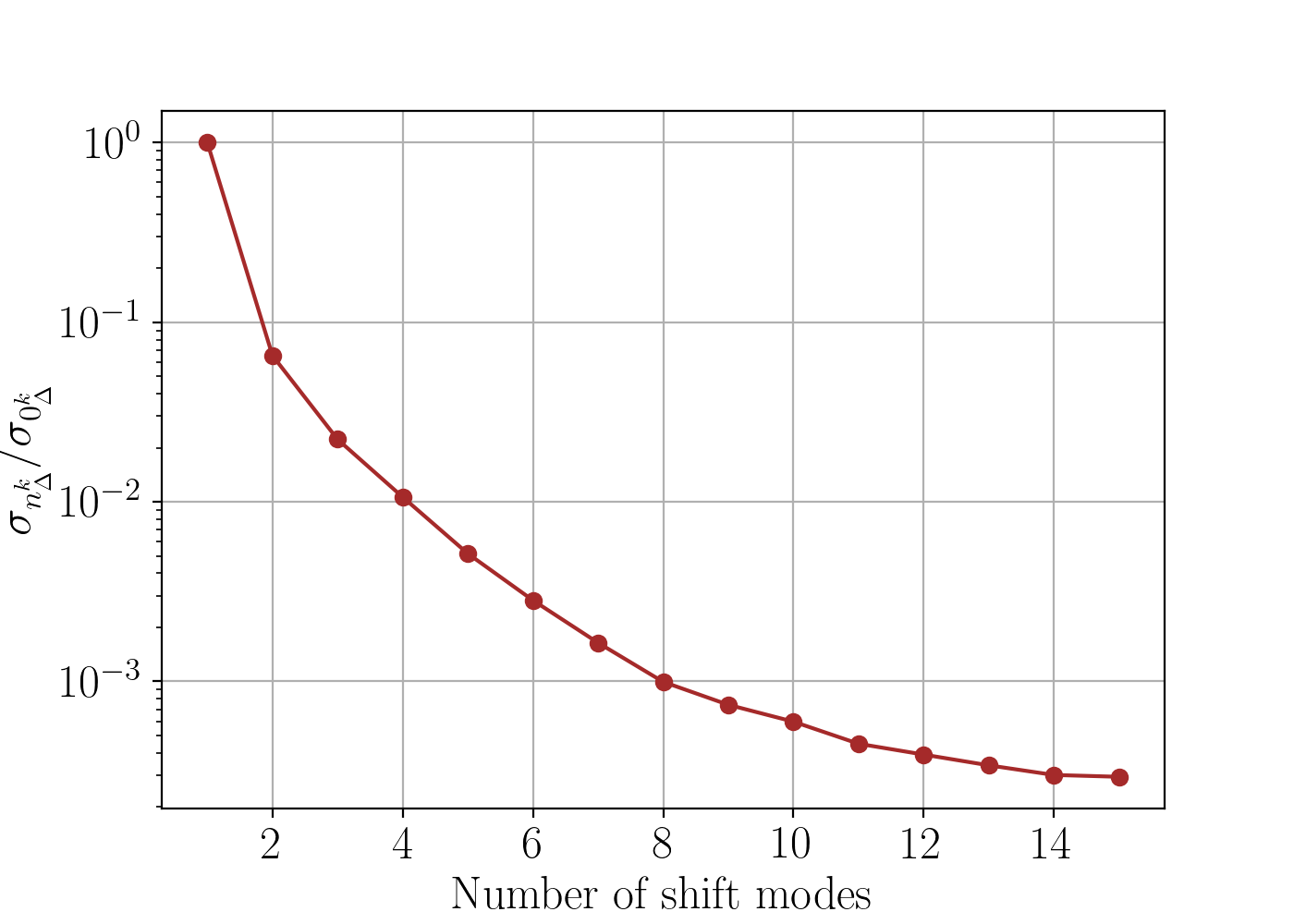}
\caption{Relative reconstruction error with respect to the number of shift truncation modes }\label{fig:shift_svd}
\end{figure}

\begin{figure}[h]
\centering
\includegraphics[scale=0.40]{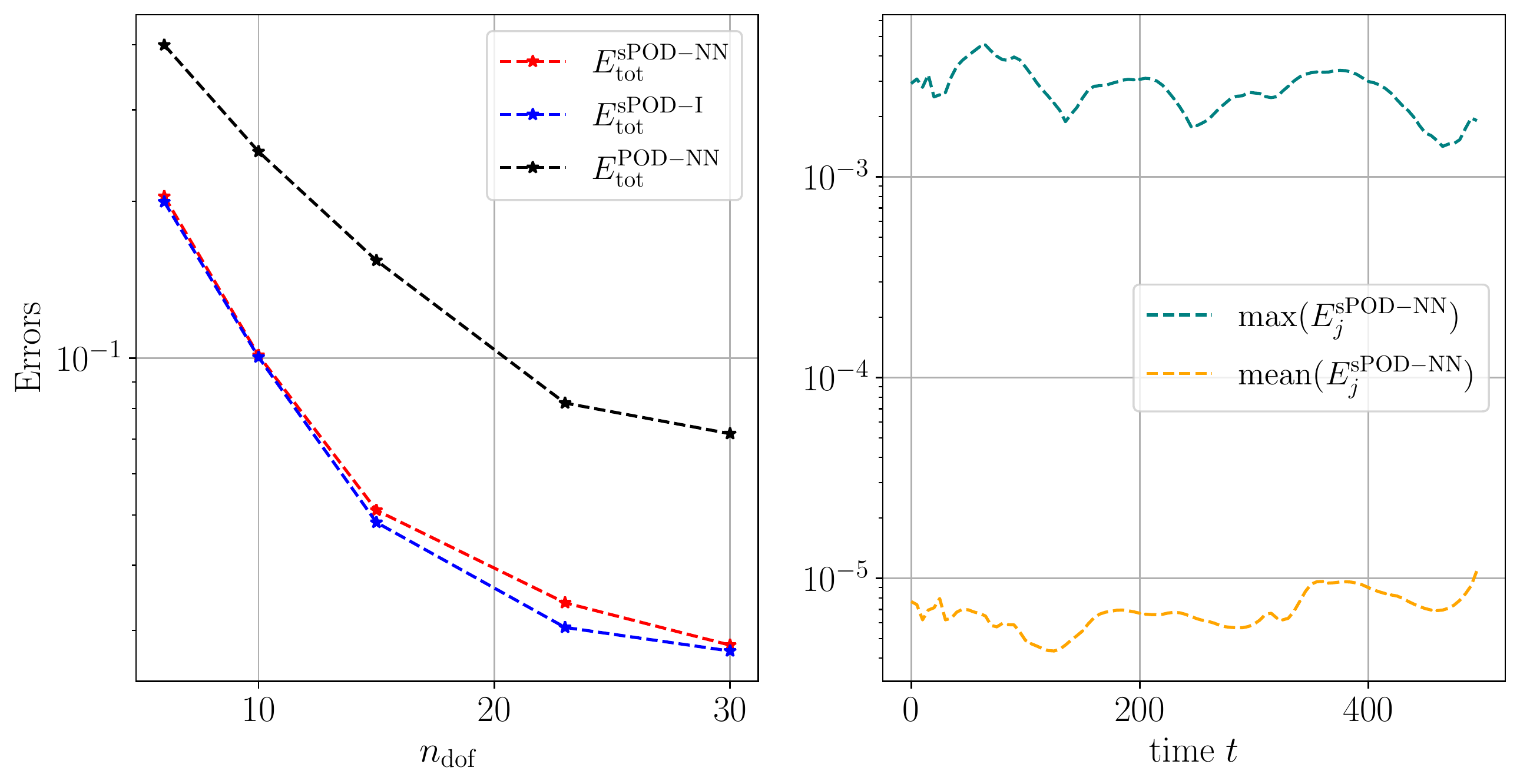}
\caption{The first plot shows the error decay for an increasing $n_{\mathrm{dof}}$. The second plot shows the trend of the relative error $E^{\mathrm{sPOD-NN}}_j$ over time.}\label{fig:wildfire2dnonlinear_full_err}
\end{figure}

\begin{figure}[h]
\centering
\includegraphics[scale=0.40]{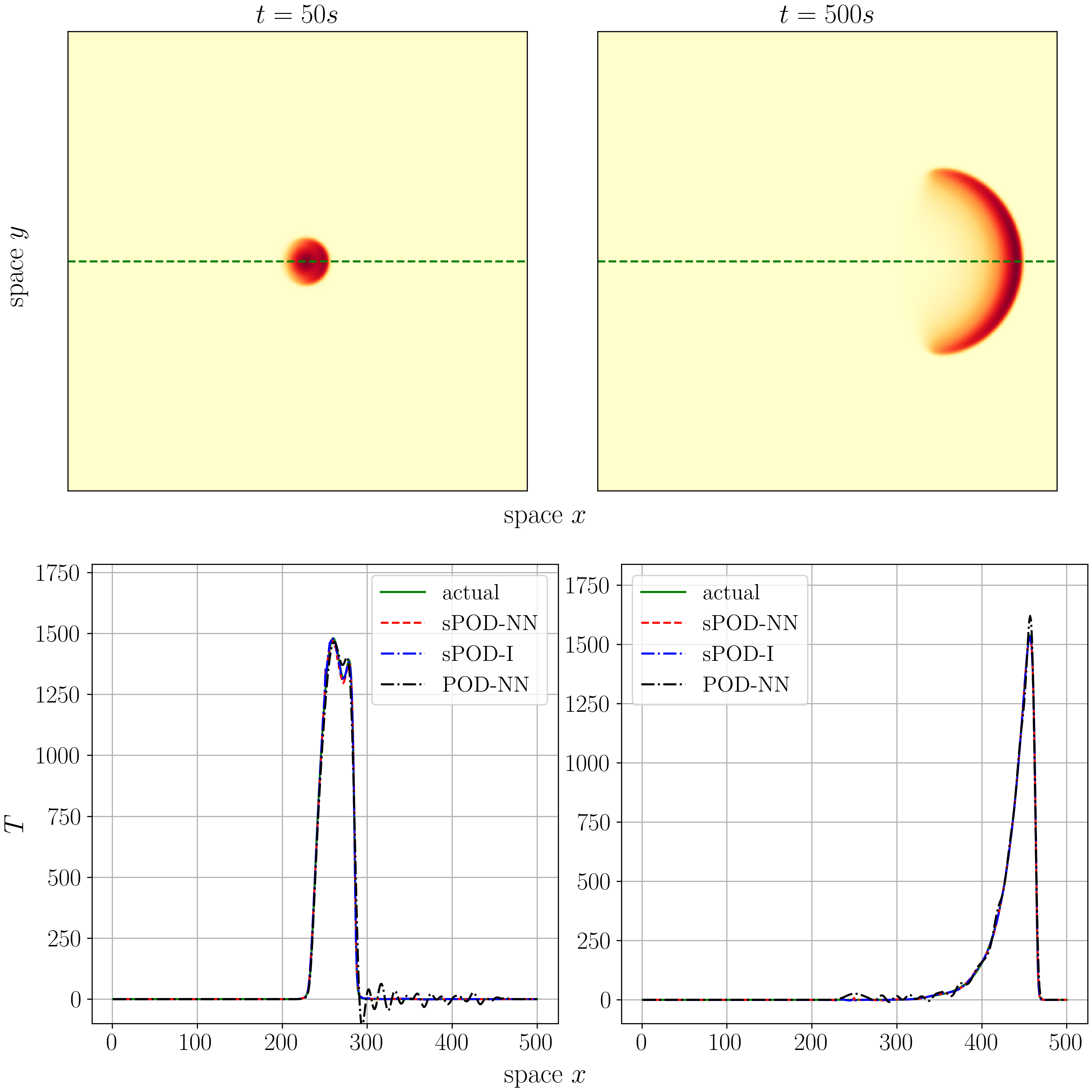}
\caption{The top two plots show the temperature profiles at $t=50s$ and $t=500s$. The bottom two plots show the cross-sectional views of the profile with the prediction results for sPOD-NN, POD-NN, and sPOD-I procedures.}\label{fig:wildfire2dnonlinear_T_x_cs}
\end{figure}

\begin{figure}[h]
\centering
\includegraphics[scale=0.3]{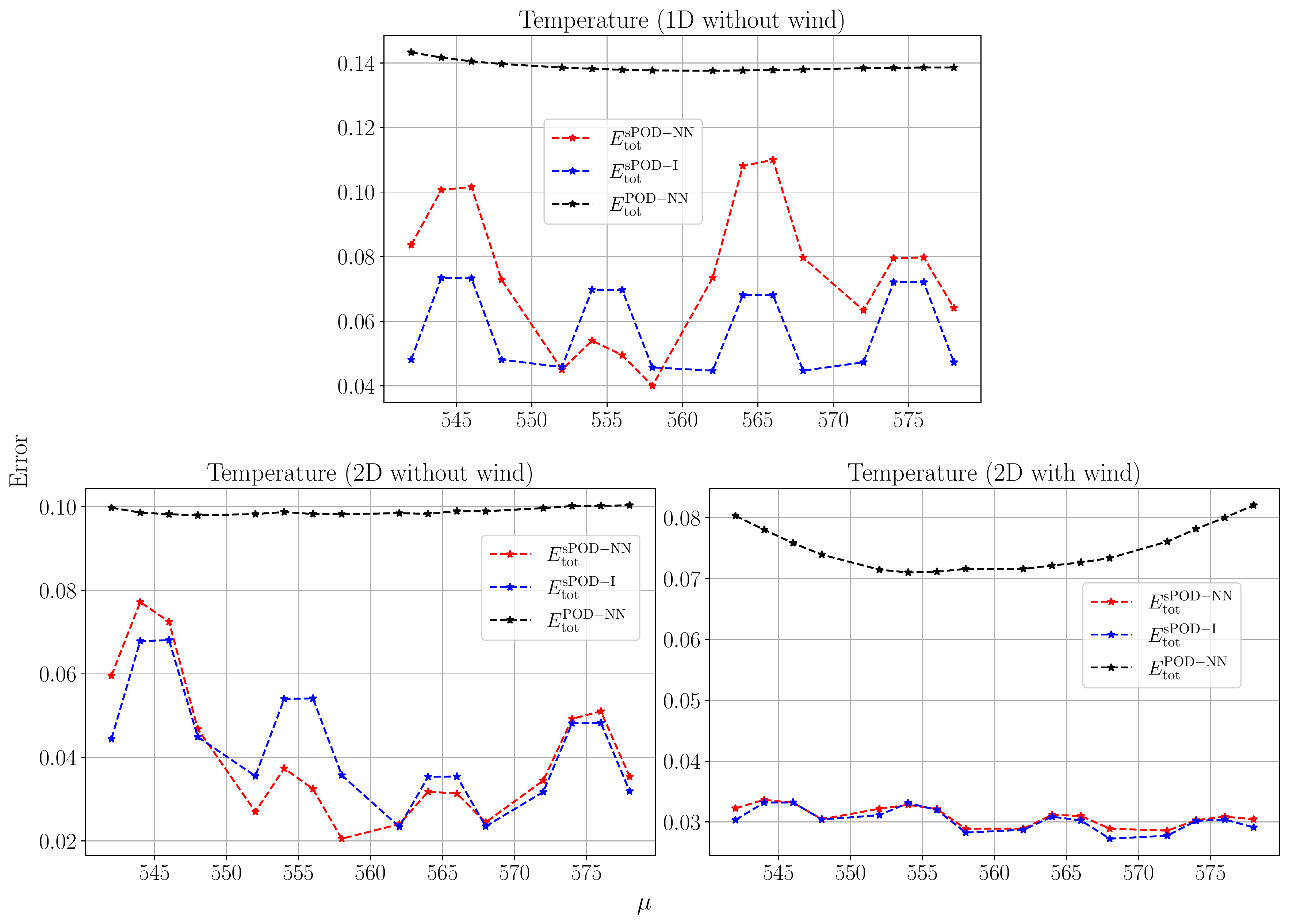}
\caption{Parameter sweep study for 1D and 2D temperature values}\label{fig:parameter_sweep_T}
\end{figure}

Non-linear parametric ROM is constructed by solving for $\mu\in\{540\mathrm{K}, 550\mathrm{K}, 560\mathrm{K}, 570\mathrm{K}, 580\mathrm{K}\}$ and sampling every $5^{th}$ snapshot in time to construct a parametric snapshot matrix $Q\in \mathbb{R}^{250000\times 500}$. The sPOD is subsequently applied for dimensionality reduction. On similar lines to the previous case, we convert our data to a polar coordinate system for simplicity. However, this time due to the changing topology of the flame front the calculation of shifts is no longer trivial. Similar to the previous case we have $\underline{\bm{\Delta}}^k = (\underline{\Delta}^k_{\mathrm{R}}, \underline{\Delta}^k_{\theta})$ and $\underline{\bm{\Delta}}^2 = (0, 0)$. However, $\underline{\Delta}^1_{\mathrm{R}}$ is now dependent on $(t, \theta, \mathrm{R})$ for which we look at \Cref{fig:wildfire2d_polar_nonlinear_cs} for detail.  
We see the edges of the fronts in the left plot for two different time instances from which we calculate the shifts for different values of $\theta$ as shown. With sPOD, we now aim at decomposing the polar temperature field into two frames: the first one captures the traveling fronts, and the second one captures the initial ignition regime. Subsequently we have $\underline{\Delta}^1_{\mathrm{R}} \in \mathbb{R}^{250000 \times N}, \: \underline{\Delta}^1_{\theta} = 0$ where in $\underline{\Delta}^1_{\mathrm{R}}$
each column gives the amount of shift to be applied to all the grid points of the domain at instances of $N$. The sPOD decomposition of the temperature is shown in \Cref{fig:wildfire2dnonlinear_sPOD_frames} in Cartesian coordinates. 
\begin{table}[h]
\begin{center}
\begin{minipage}{\textwidth}
\caption{Offline and online error study w.r.t number of modes for 2D wildland fire model (with wind)}\label{tab:2dwildfirenonlinear_errors}
\begin{tabular*}{\textwidth}{@{\extracolsep{\fill}}lcccccc@{\extracolsep{\fill}}}
\toprule%
Modes\footnotemark[1]& \multicolumn{2}{@{}c@{}}{Offline errors} & \multicolumn{3}{@{}c@{}}{Online errors} \\\cmidrule{2-3}\cmidrule{4-6}%
$r_1+r_2$ & $E^{\mathrm{sPOD}}$ & $E^{\mathrm{POD}}$ & $E^{\mathrm{sPOD-NN}}_{\mathrm{tot}}$ & $E^{\mathrm{sPOD-I}}_{\mathrm{tot}}$ & $E^{\mathrm{POD-NN}}_{\mathrm{tot}}$ \\
\midrule
1 + 1 & $2.63\times 10^{-1}$ & $4.01\times 10^{-1}$ & $2.04\times 10^{-1}$ & $1.99\times 10^{-1}$ & $3.99\times 10^{-1}$ \\
4 + 2 & $9.88\times 10^{-2}$ & $2.51\times 10^{-1}$ & $1.01\times 10^{-1}$ & $1.00\times 10^{-1}$ & $2.49\times 10^{-1}$ \\
8 + 3 & $4.73\times 10^{-2}$ & $1.55\times 10^{-1}$ & $5.10\times 10^{-2}$ & $4.83\times 10^{-2}$ & $1.54\times 10^{-1}$ \\
13 + 6 & $2.46\times 10^{-2}$ & $8.45\times 10^{-2}$ & $3.39\times 10^{-2}$ & $3.04\times 10^{-2}$ & $8.19\times 10^{-2}$ \\
17 + 9 & $1.97\times 10^{-2}$ & $5.75\times 10^{-2}$ & $2.81\times 10^{-2}$ & $2.73\times 10^{-2}$ & $7.16\times 10^{-2}$ \\
\botrule
\end{tabular*}
\footnotetext[1]{We also have $r=r_1 + r_2 + 4$, where $r_1$ is the rank of the moving frame, $r_2$ is the rank of the stationary frame, and $4$ accounts for the added number of amplitudes extracted from the shift matrix $\underline{\bm{\Delta}}^1_{\mathrm{R}}$. For comparison, we therefore use $r$ modes for the POD.}
\end{minipage}
\end{center}
\end{table}
The basis reconstruction error results are shown in \Cref{tab:2dwildfirenonlinear_errors}. We could reach to $E^{\mathrm{sPOD}}\sim 0.01$ with $17+9=26$ modes whereas $E^{\mathrm{POD}}\sim 0.06$.

sPOD gives us the time amplitudes for both the frames and we proceed to construct the time amplitude matrix $\hat{A}$ for training as explained in \Cref{sec:theory}. The shift truncation dimension $\shiftDim^1 = 4$ is chosen suitably keeping in mind the relative reconstruction error of the shift matrix shown in \Cref{fig:shift_svd}. This is one particular case of extracting low-dimensional structure from the high-dimensional shifts which is desired to proceed with the neural network training. The theory behind this is explained in detail in \Cref{sec:theory}.
For this problem $\hat{A} \in \mathbb{R}^{30 \times 500}$ and $P \in \mathbb{R}^{2 \times 500}$. The network parameters remain the same as the previous case with $N_{\mathrm{epochs}}=200000$ and batch size $N_b=50$. Early stopping criteria are imposed to prevent overfitting after 4000 consecutive epochs. For the test parameter, $\mu = [558.49\mathrm{K}]$ the time amplitude and the shift predictions are in line with the previous case. We also show the results of a parameter sweep study for the temperature for 16 different test samples where we sample the values for $\mu\in [540, 580]$. The result is shown in \Cref{fig:parameter_sweep_T}. Error estimates are shown in \Cref{fig:wildfire2dnonlinear_full_err}. In the first plot we observe the $E^{\mathrm{sPOD-NN}}_{\mathrm{tot}} \sim 0.028$. In the second plot, we see the error over the whole time interval $E^{\mathrm{sPOD-NN}}_j$ where we see $\max < \mathcal{O}(10^{-2})$.
Full reconstruction results are shown in \Cref{fig:wildfire2dnonlinear_T_x_cs}. The cross-sectional views of the temperature profile at two different time instances are shown. We observe that the POD-NN is corrupted by oscillations but for sPOD-NN and sPOD-I we do not see any oscillations.

\subsection{Timing study}
\begin{figure}[h]
\centering
\includegraphics[scale=0.43]{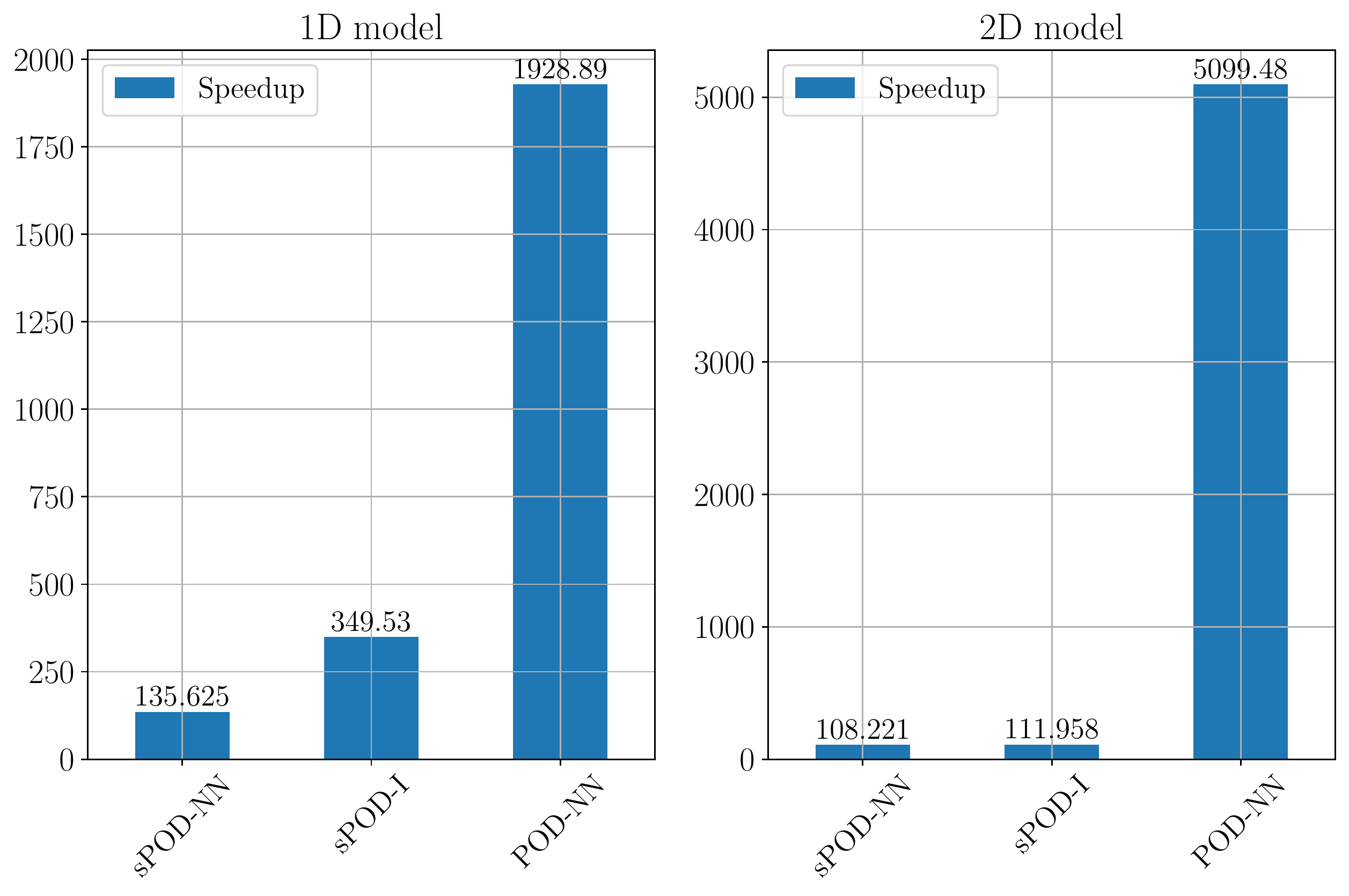}
\caption{Computational time analysis for the online phase. For 1D model $n_{\mathrm{dof}} = 44$, for 2D model (without wind) $n_{\mathrm{dof}} = 25$ and for 2D model (with wind) $n_{\mathrm{dof}} = 30$. The $t_{\mathrm{FOM}} = 5.2 s$ for 1D model and $t_{\mathrm{FOM}} = 63.2 s$ for the 2D model.}\label{fig:timing}
\end{figure}

Here we show the computational time analysis of the proposed methods. All the tests for computing the timing are run on Macbook Air M1(2020) with an 8-core CPU and 16GB of RAM. We refer to \Cref{fig:timing} where we see that for both 1D and 2D models, both sPOD-NN and sPOD-I methods are more than 100 times faster than the FOM for the multi-query scenario. However, this speedup becomes very large (more than 1000 times) for POD-NN. The reason for this could be explained by looking at the sub-steps of sPOD-NN and POD-NN. For the 1D model, the sPOD-NN timing consists of the time for evaluation of the neural network followed by that of transforming the stationary frames into co-moving frames which involves matrix-matrix multiplications for every frame and at last the time for adding the individual frames and producing the final snapshot. However, for POD-NN the timing only consists of evaluating the neural network and reconstructing the final snapshot with just one matrix-matrix multiplication operation. This difference translates into the 2D model as well except we have one more sub-step for sPOD-NN where the snapshot data has to be converted from a Cartesian coordinate system to a polar one for analysis and back again. This makes the difference in speedups more pronounced. The speedup shown in \Cref{fig:timing} is defined as:
\begin{equation}
    \text{Speedup} = \frac{t_{\mathrm{FOM}}}{t_{\mathrm{ROM}}}
\end{equation}
where $t_{\mathrm{ROM}}$ is the time taken for the methods proposed in the paper. The $t_{\mathrm{FOM}} = 5.2 s$ for 1D model and $63.2 s$ for the 2D model. Both the test cases for the 2D model described in the paper have similar computational time consumption. The comparison is made with the same $n_{\mathrm{dof}}$ across all methods for a specific model. With the results shown, we substantiate our claims that not only is the proposed method accurate but is also extremely fast for the online phase. 
Just for the sake of completeness, the time study for the offline phase is shown in \Cref{tab:timing_offline}.
\begin{table}[h]
\begin{center}
\begin{minipage}{\textwidth}
\caption{Offline phase timings}\label{tab:timing_offline}
\begin{tabular*}{\textwidth}{@{\extracolsep{\fill}}lcccccc@{\extracolsep{\fill}}}
\toprule%
& \multicolumn{2}{@{}c@{}}{Basis reconstruction time} & \multicolumn{2}{@{}c@{}}{Network training time} \\\cmidrule{2-3}\cmidrule{4-5}%
Model & sPOD & POD & sPOD-NN & POD-NN \\
\midrule
1D & $\sim 1.5$ h & $54$ s & $\sim 32$ min & $\sim 32$ min \\
2D (without wind) & $\sim 33$ h & $34$ s & $\sim 9$ min & $\sim 9$ min \\
2D (with wind) & $\sim 12$ min & $34$ s & $\sim 9$ min & $\sim 9$ min \\
\botrule
\end{tabular*}
\end{minipage}
\end{center}
\end{table}

\section{Conclusion}\label{sec:conclusion}
In this paper, we proposed a non-intrusive model reduction technique sPOD-NN for parametric transport-dominated systems. In particular, the technique uses sPOD for constructing the reduced basis and then subsequently extracts the time amplitudes for training a neural network for the offline phase. In the online phase, the trained model is then used to predict the time amplitudes at unseen parameter values. The core concept builds upon the ability of sPOD, over the conventional methods like POD in constructing an optimal reduced basis for transport-dominated problems. For benchmarking we tested sPOD-NN against two other methods: sPOD-I and POD-NN. We assessed the computational performance and the prediction accuracy of the methods on non-linear time-dependent parametrized PDE systems: 1D and 2D wildland fire models. Through the numerical results shown in \Cref{sec:results}, we saw that sPOD-NN yielded accurate numerical approximations for the time amplitudes and the shifts and in turn the final snapshot. Along with the accurate predictions the proposed method provides substantial speedups in the online phase compared to the FOM. We also saw that the sPOD-NN outperforms POD-NN in all the examples presented owing to the ability of sPOD to better construct the optimal reduced basis which will almost always happen in transport-dominated systems. Also, based on the results we saw that the scope of improvement for sPOD-NN is huge compared to POD-NN as the total error is dominated by the neural network prediction error for sPOD-NN. Given more number of training samples the network prediction error could be further reduced. 

For the scope of this paper, we considered models with either no wind presence or a constant unidirectional wind. Although for such test cases, the presented methods would suffice, they will have to be modified to handle more realistic, non-uniform, and shape-changing fronts \cite{krah_front_2022}. Also, the wildland fire model considered here although is enough to capture the overall dynamics of the process broadly, more complex and detailed models could be considered as test cases for the proposed methods. As a future step, our aim is to study a more challenging test case where the front profiles change in a more complicated way as time progresses. In such cases, it becomes even more difficult to accurately compute the shifts from the snapshot data making it a perfect candidate to run our proposed methods on and substantiate their generality.

\section*{Code and Data Availability}

In order to facilitate reproducibility and transparency of the research presented in this paper, the source code used for the experiments and analyses is made publicly available. Interested parties can access the code at the following GitHub repository:\\
\begin{center}
\urlstyle{tt}
\url{https://github.com/MOR-transport/sPOD-NN-paper}
\end{center}
\vspace{1em}
We encourage researchers to make use of the code and to extend it for their own research purposes. In addition, the trained network-data is available upon request, and interested parties can contact the corresponding author for access.

\addcontentsline{toc}{section}{Author Contribution Statement (CRediT)}
\section*{Author Contribution Statement (CRediT)}

In the following, we declare the authors' contributions to this work.

\vspace{5pt}
{\small
\noindent
\begin{tabular}{@{}p{0.3\linewidth} p{0.7\linewidth}}
\textbf{Shubhaditya Burela:} & methodology, implementation of neural networks, computations, and visualizations, writing original draft, reviewing \& editing\\
\textbf{Philipp Krah:} & initial concept, methodology, implementation of sPOD algorithm, reviewing \& editing\\
\textbf{Julius Reiss:} &  supervision, reviewing \& editing
\end{tabular}
}

\bibliography{sn-bibliography}% common bib file


\appendix
\section*{Appendix}

\section{Neural network foundation}\label{apx:NNarchitecture}
In recent years, non-intrusive model reduction methods that utilize neural networks have gained considerable attention. Despite advancements in the type and architecture of these networks, the core concept remains unchanged. In this section, we describe the specific neural network architecture used in the numerical examples presented in this work.

\begin{figure}[h]
\centering
\includegraphics[scale=0.38]{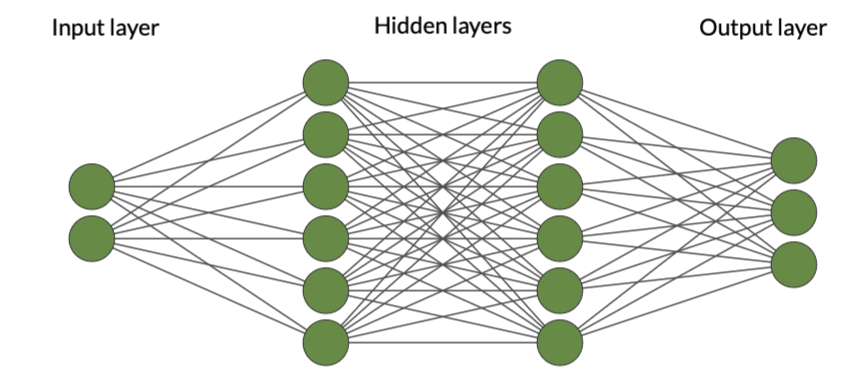}
\caption{Feed-forward neural network}\label{fig:app_network}
\end{figure}

For the architecture, we use a deep FNN (\Cref{fig:app_network}). For a clearer picture, consider a feed-forward network with $L$ hidden layers where $l\in \{1, \ldots, L\}$ is the index of the hidden layer, for such a setup the forward propagation is described as:
\begin{align}\label{eq:app_network_forw}
\begin{split}
 z^{(l+1)}_i &= \bm{w}^{(l+1)}_i \bm{y}^{(l)} + b^{(l+1)}_i ,
\\
 y^{(l+1)}_i &= f(z^{(l+1)}_i)
\end{split}
\end{align}
where $\bm{y}^{(l)}$ is the vector of outputs from the layer $l$, $\bm{z}^{(l)}$ is the vector of inputs into the layer $l$, $W^{(l)}$ is the weight matrix in the layer $l$ along with a bias vector for the layer $l$, $\bm{b}^{(l)}$. The activation function is given by $f(\cdot)$. In the backpropagation step, the training of the network is usually carried out with a gradient descent algorithm minimizing a certain loss function $J(W, \bm{b})$. As we deal with a regression problem in our work we basically rely on two types of loss functions, namely MAE (Mean Absolute Error) often referred to as L1 loss, and MSE (Mean Squared Error):
\begin{equation}
    MAE = \frac{1}{n}\sum^{n}_{i=1}\left \lvert y_i - \hat{y}_i\right \rvert, \qquad MSE = \frac{1}{n}\sum^{n}_{i=1}(y_i - \hat{y}_i)^2
\end{equation}
where $y_i$ and $\hat{y}_i$ are the input and the target vectors respectively. The training data in our problems is heavily skewed and contains critical outliers that impact the physical significance of the problem. Using Mean Squared Error (MSE) as a loss function can penalize these outliers and produce subpar prediction results. To address this, we use Mean Absolute Error (MAE) as the preferred loss function due to its robustness in the presence of outliers. A stochastic gradient descent algorithm with momentum term is utilized for minimizing the loss function and updating the parameters $W$ and $\bm{b}$ in every iteration as shown:
\begin{align}\label{eq:app_network_back}
\begin{split}
 &W^{(l)}_{ij} \leftarrow W^{(l)}_{ij} - \alpha \frac{\partial}{\partial W^{(l)}_{ij}} J(W, \bm{b}) ,
\\
 &b^{(l)}_{i} \leftarrow b^{(l)}_{i} - \alpha \frac{\partial}{\partial b^{(l)}_{i}} J(W, \bm{b})
\end{split}
\end{align}
$\alpha$ is the learning rate and can be tuned to get a better decay of the loss function. $W^{(l)}_{ij}$ is an element of the weight matrix associated with the connection between a node $j$ in layer $l$ and a node $i$ in layer $l+1$ and the term $b^{(l)}_i$ is an element of the bias vector associated with the node $i$ in layer $l+1$. 
The \Cref{tab:app_network_params} outlines the activation functions used. The first three layers use Exponential Linear Unit (ELU) activation, and the next layer uses Leaky Rectified Linear Unit (LeakyReLU). It is common in the literature to not use any activation function for the output layer in regression problems, which we also follow in this case. The detailed comparison of the different activation functions can be found in \cite{jagtap_how_2022}, \cite{nwankpa_activation_2018}.

\section{Results for supply mass fraction}\label{apx:results_supply_mass_fraction}
\begin{figure}[h]
\centering
\includegraphics[scale=0.3]{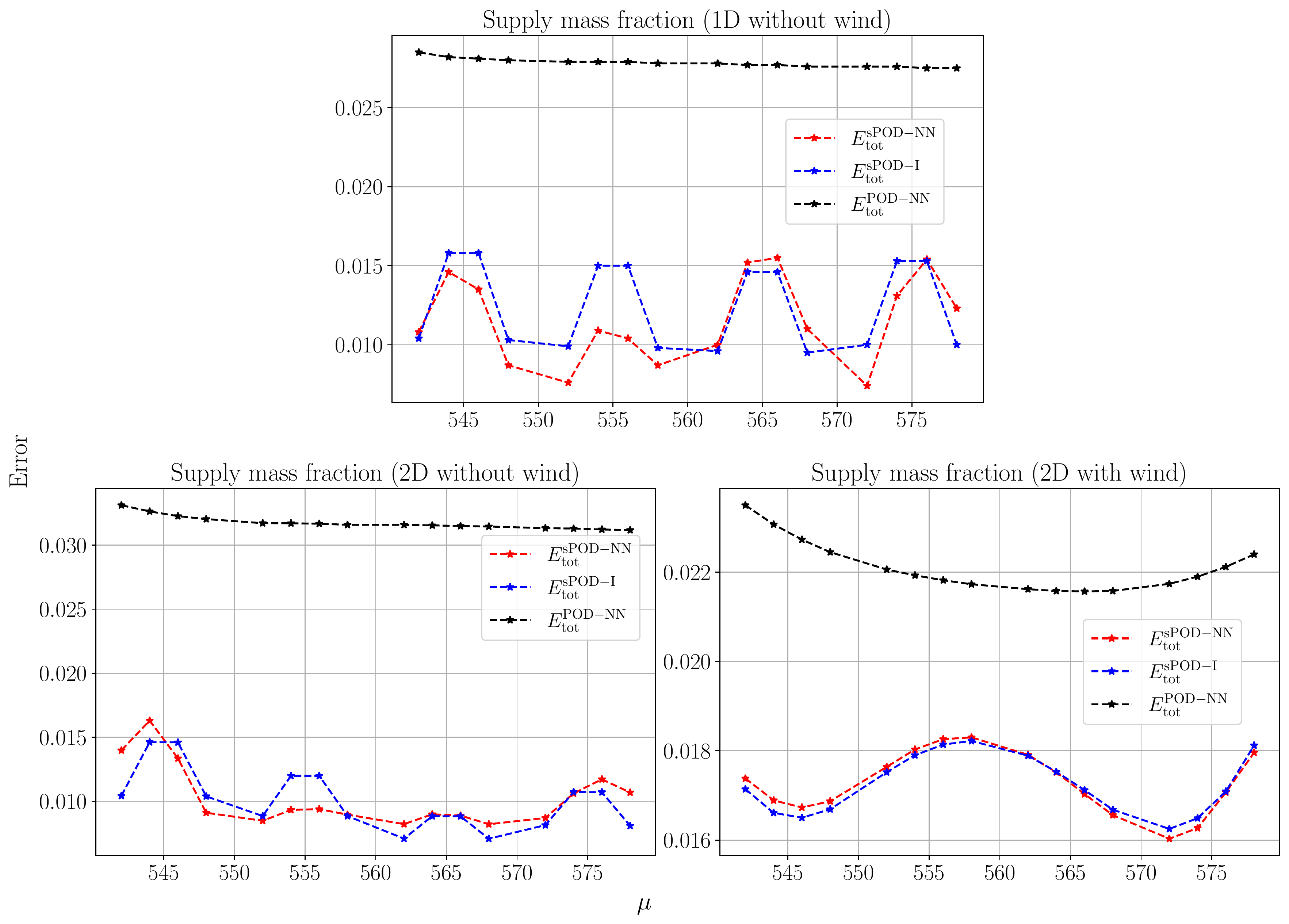}
\caption{Parameter sweep study for 1D and 2D supply mass fraction values}\label{fig:parameter_sweep_S}
\end{figure}

Here we present the results for the supply mass fraction for all the models presented in the text. \Cref{tab:1dwildfire_errors_S}, \Cref{tab:2dwildfire_errors_S} and  \Cref{tab:2dwildfirenonlinear_errors_S} show the offline and online errors for 1D, 2D(without wind) and 2D(with wind) models. \Cref{fig:parameter_sweep_S} shows the parameter sweep study for supply mass fraction where we test the model on the samples of $\mu\in [540, 580]$ 
\begin{table}[h]
\begin{center}
\begin{minipage}{\textwidth}
\caption{Offline and online error study w.r.t number of modes for 1D wildland fire model (supply mass fraction)}\label{tab:1dwildfire_errors_S}
\begin{tabular*}{\textwidth}{@{\extracolsep{\fill}}lcccccc@{\extracolsep{\fill}}}
\toprule%
& \multicolumn{2}{@{}c@{}}{Offline errors} & \multicolumn{3}{@{}c@{}}{Online errors} \\\cmidrule{2-3}\cmidrule{4-6}%
Modes & $E^{\mathrm{sPOD}}$ & $E^{\mathrm{POD}}$ & $E^{\mathrm{sPOD-NN}}_{\mathrm{tot}}$ & $E^{\mathrm{sPOD-I}}_{\mathrm{tot}}$ & $E^{\mathrm{POD-NN}}_{\mathrm{tot}}$ \\
\midrule
1 + 1 + 1 & $8.95\times 10^{-2}$ & $1.53\times 10^{-1}$ & $6.27\times 10^{-2}$ & $6.17\times 10^{-2}$ & $1.53\times 10^{-1}$ \\
4 + 2 + 4 & $2.15\times 10^{-2}$ & $6.76\times 10^{-2}$ & $2.19\times 10^{-2}$ & $1.79\times 10^{-2}$ & $6.70\times 10^{-2}$ \\
8 + 2 + 8 & $3.88\times 10^{-3}$ & $4.05\times 10^{-2}$ & $8.91\times 10^{-3}$ & $8.62\times 10^{-3}$ & $4.06\times 10^{-2}$ \\
12 + 4 + 12 & $9.00\times 10^{-4}$ & $2.44\times 10^{-2}$ & $7.35\times 10^{-3}$ & $7.81\times 10^{-3}$ & $2.78\times 10^{-2}$ \\
\botrule
\end{tabular*}
\end{minipage}
\end{center}
\end{table}
\begin{table}[h]
\begin{center}
\begin{minipage}{\textwidth}
\caption{Offline and online error study w.r.t number of modes for 2D (without wind) wildland fire model (supply mass fraction)}\label{tab:2dwildfire_errors_S}
\begin{tabular*}{\textwidth}{@{\extracolsep{\fill}}lcccccc@{\extracolsep{\fill}}}
\toprule%
& \multicolumn{2}{@{}c@{}}{Offline errors} & \multicolumn{3}{@{}c@{}}{Online errors} \\\cmidrule{2-3}\cmidrule{4-6}%
Modes & $E^{\mathrm{sPOD}}$ & $E^{\mathrm{POD}}$ & $E^{\mathrm{sPOD-NN}}_{\mathrm{tot}}$ & $E^{\mathrm{sPOD-I}}_{\mathrm{tot}}$ & $E^{\mathrm{POD-NN}}_{\mathrm{tot}}$ \\
\midrule
1 + 1 & $9.60\times 10^{-2}$ & $1.24\times 10^{-1}$ & $7.71\times 10^{-2}$ & $7.79\times 10^{-2}$ & $1.24\times 10^{-1}$ \\
4 + 1 & $4.20\times 10^{-2}$ & $7.59\times 10^{-2}$ & $4.17\times 10^{-2}$ & $4.14\times 10^{-2}$ & $7.51\times 10^{-2}$ \\
7 + 3 & $7.45\times 10^{-3}$ & $4.57\times 10^{-2}$ & $1.04\times 10^{-2}$ & $9.06\times 10^{-3}$ & $4.57\times 10^{-2}$ \\
11 + 4 & $5.39\times 10^{-3}$ & $3.10\times 10^{-2}$ & $8.73\times 10^{-3}$ & $7.77\times 10^{-3}$ & $3.15\times 10^{-2}$ \\
\botrule
\end{tabular*}
\end{minipage}
\end{center}
\end{table} 
\begin{table}[h]
\begin{center}
\begin{minipage}{\textwidth}
\caption{Offline and online error study w.r.t number of modes for 2D (with wind) wildland fire model (supply mass fraction)}\label{tab:2dwildfirenonlinear_errors_S}
\begin{tabular*}{\textwidth}{@{\extracolsep{\fill}}lcccccc@{\extracolsep{\fill}}}
\toprule%
& \multicolumn{2}{@{}c@{}}{Offline errors} & \multicolumn{3}{@{}c@{}}{Online errors} \\\cmidrule{2-3}\cmidrule{4-6}%
Modes & $E^{\mathrm{sPOD}}$ & $E^{\mathrm{POD}}$ & $E^{\mathrm{sPOD-NN}}_{\mathrm{tot}}$ & $E^{\mathrm{sPOD-I}}_{\mathrm{tot}}$ & $E^{\mathrm{POD-NN}}_{\mathrm{tot}}$ \\
\midrule
1 + 1 & $7.43\times 10^{-2}$ & $3.67\times 10^{-2}$ & $7.15\times 10^{-2}$ & $7.13\times 10^{-2}$ & $3.53\times 10^{-2}$ \\
2 + 2 & $2.95\times 10^{-2}$ & $2.92\times 10^{-2}$ & $2.09\times 10^{-2}$ & $2.08\times 10^{-2}$ & $2.73\times 10^{-2}$ \\
4 + 3 & $1.71\times 10^{-2}$ & $2.23\times 10^{-2}$ & $1.82\times 10^{-2}$ & $1.82\times 10^{-2}$ & $2.17\times 10^{-2}$ \\
\botrule
\end{tabular*}
\end{minipage}
\end{center}
\end{table} 

\section{sPOD parameters}\label{apx:sPOD_parameters}
As suggested by \cite{black_efficient_2021} we use the sPOD for decomposing the snapshot data of the wildland fire model. To determine the co-moving fields $\{\matr{Q}^k\}_{k=1,\dots,f}$ shown in \Cref{eq-def:sPOD-decomposition-discrete} and an error term $E\in\mathbb{R}^{M\times N}$ to capture noise, we solve the following constraint optimization problem:
\begin{align}
    \label{opt:shifted-robust-PCA}
        \min_{\matr{Q}^k,\matr{E}} \sum_{k=1}^f\norm{\matr{Q}^k}_* + \varsigma\norm{\matr{E}}_1
        \quad\text{s.t. }\quad
        \matr{Q} = \sum_{k=1}^{f} T^{\Delta^k} (\matr{Q}^k) + \matr{E}\,,
\end{align}
using algorithm 8 in \cite{krah_non-linear_nodate}.
Here, $\Vert E \Vert_1 = \sum_{ij} \lvert E_{ij} \rvert$ is not the usual matrix 1-norm, but the vector 1-norm of a long vector $E\in\mathbb{R}^{M\times N}$, $\Vert A \Vert_* = \sum_i \sigma_i(A)$ is the Schatten one-norm and $\varsigma\ge0,\eta>0$ are tuning parameters. For the 1D and 2D wildland fire model, we state $\varsigma,\eta$ in \Cref{tab:sPOD_params}.
It is important to note that once the computation of ${\matr{Q}^k}_{k=1,\dots,f}$ and $E\in\mathbb{R}^{M\times N}$ is complete, $E\in\mathbb{R}^{M\times N}$ is disregarded as it solely captures the data noise. 

\begin{table}[h!]
\begin{center}
\begin{minipage}{\textwidth}
\caption{sPOD parameters for 1D and 2D wildland fire models}
\label{tab:sPOD_params}
\centering
\begin{tabular}{c c c c} 
 \toprule
 Model & $\eta$ & $\varsigma$ & $N_{\mathrm{iter}}$ \\
 \midrule
 1D & $5.47\times 10^{-6}$ & $1.85\times 10^{-1}$ & $2000$ \\ 

 2D (without wind) & $3.92\times 10^{-7}$ & $4.47\times 10^{-2}$ & $2000$ \\

 2D (with wind) & $1.18\times 10^{-5}$ & $4.47\times 10^{2}$ & $7$ \\
 \bottomrule
\end{tabular}
\end{minipage}
\end{center}
\end{table}

\section{Synthetic test case}\label{apx:synthetic_test_case}
\begin{figure}[h]
\centering
\includegraphics[scale=0.5]{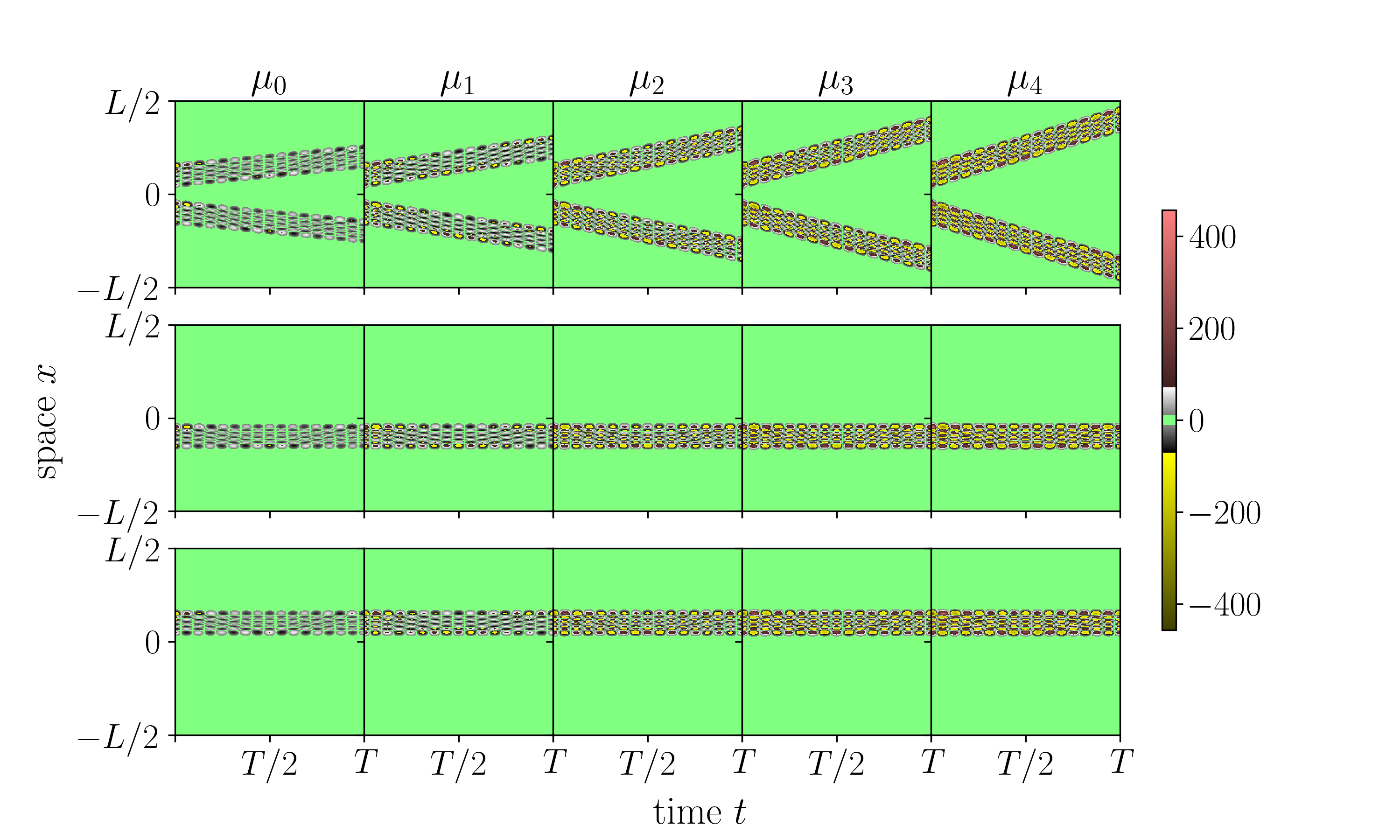}
\caption{The second and the third row of the plot shows $q^1$ and $q^2$ respectively which are the individual waves and the top plot shows $q$ which is the shifted superposed solution computed at $\mu = [0.1, 0.15, 0.2, 0.25, 0.3]$ respectively.}\label{fig:synthetic_data}
\end{figure}

\begin{figure}[h]
\centering
\includegraphics[scale=0.5]{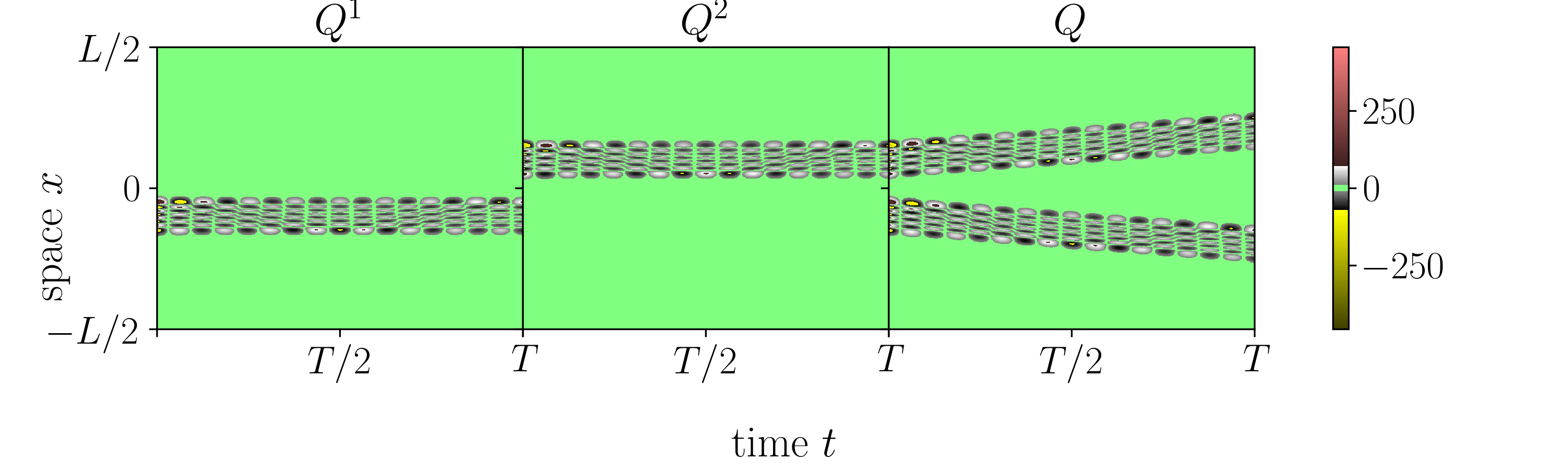}
\caption{sPOD decomposition on $Q$ corresponding to $\mu_0 = 0.1$. The first two images are the decomposed stationary frames and the last image is the reconstructed $Q$.}\label{fig:synthetic_spod_frames}
\end{figure}

\begin{figure}[h]
\centering
\includegraphics[scale=0.38]{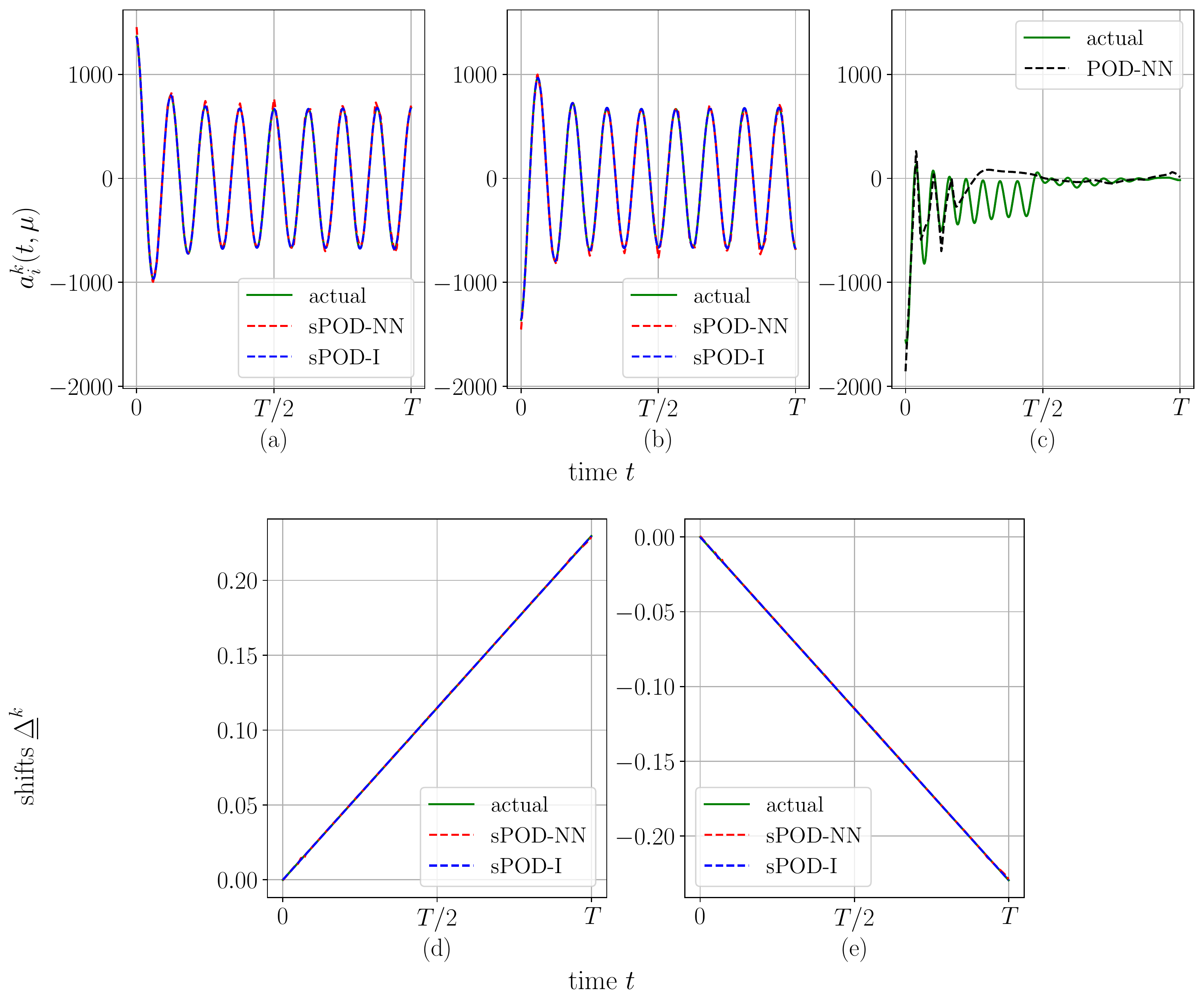}
\caption{(a) and (b) show the prediction results for the first mode of both the frames with sPOD-NN and sPOD-I. (c) is the prediction result for the first mode with the POD-NN approach. (d) and (e) show the prediction results for the shifts for both frames.}\label{fig:synthetic_time_amplitudes_pred}
\end{figure}

\begin{figure}[h]
\centering
\includegraphics[scale=0.40]{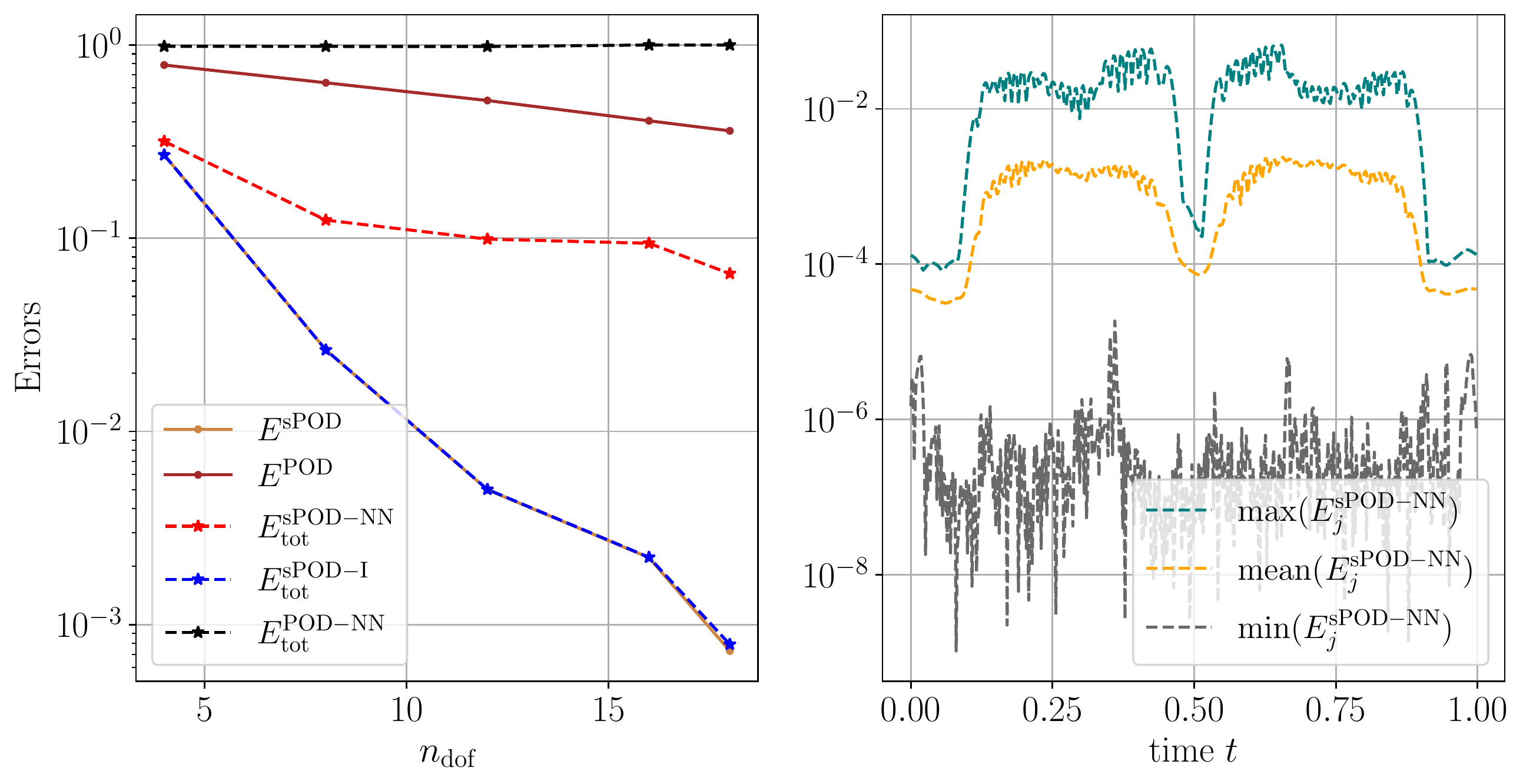}
\caption{The first plot shows the error decay for an increasing $n_{\mathrm{dof}}$. $E^{\mathrm{sPOD}}$, $E^{\mathrm{POD}}$ show the basis reconstruction error for sPOD and POD respectively, and the errors labeled $E_{\mathrm{tot}}$ show the final reconstruction error after prediction or interpolation. The second plot shows the trend of the relative error $E^{\mathrm{sPOD-NN}}_j$ over time.}\label{fig:synthetic_full_err}
\end{figure}

As a proof of concept for the proposed method, we consider a 1D test case of traveling waves $q^1$ and $q^2$. We assume that both waves move along the path:
\begin{equation}
    \underline{\bm{\Delta}}^1(t, \mu) = \mu t, \quad \quad \text{and} \quad \quad \underline{\bm{\Delta}}^2(t, \mu) = -\mu t
\end{equation}
The path is parameterized by the parameter $\mu$. We have $\mu \in (\mu_1, \ldots, \mu_{N_p}) \in \mathcal{P} \subset \mathbb{R}^{N_p}$ which means that for different values of the parameter $\mu$ the paths of the traveling waves will change. We then assume a function $q(x, t,\mu)$ which is constructed by the superposition of $q^1$ and $q^2$:
\begin{equation}
    q(x, t,\mu) = q^1(x + \underline{\bm{\Delta}}^1(t, \mu), t,\mu) + q^2(x + \underline{\bm{\Delta}}^2(t, \mu), t,\mu),
\end{equation}
where $(x, t, \mu) \in [-L/2, L/2[\times[0, T[\times \mathcal{P}$, with $M$ number of grid points in the spatial domain, $N_t$ time steps and $N_p$ instances of the parameter $\mu$. To be as close as possible to a realistic setting, we assume that $q(x, t,\mu)$ is the solution of a high dimensional ODE resulting from a discretized PDE. We also assume that we only have access to the snapshot matrix $Q_{i, j + N_t*p} = q(x_i, t_j, \mu_p)$ and the paths $\underline{\bm{\Delta}}^k(t_j)$ for $N_p$ parameters and where $k$ is the number of individual waves being superposed to construct the solution. The individual waves for a single parameter instance $\mu_p$ are defined as:
\begin{multline}\label{eq_def:syndata}
    q^k(x, t,\mu_p) := \sum^{n_s-1}_{n=0} \mu_p\left(1 + e^{-2nt}\right) \cos\left(-2\pi t/T(n+1)\right)h_n(x^k/w), \\
    \langle h_n, h_m \rangle = \delta_{n, m}
\end{multline}
where $w=0.015L$ and $x^k = x \pm 0.1L$ for $k=1,2$ respectively. Both $q^1$ and $q^2$ are defined in terms of dyadic pairs as shown in \Cref{eq_def:syndata} as it enables us to tune the singular value spectra ($n_s$ being the number of singular values) in each frame. Furthermore, we choose Gauss-Hermite polynomials $h_n$
\begin{equation}
    h_n(x) = \frac{(-1)^n}{\sqrt{2^n n!\sqrt{\pi}}}e^{x^2/2}\frac{d^n}{dx^n}e^{-x^2} = \frac{1}{\sqrt{2^n n!\sqrt{\pi}}} H_n(x)e^{-\frac{1}{2}x^2}
\end{equation}
as they nicely mimic strongly localized wave structures (since $w \ll L$). To build the data, we consider $M=500, \: N_t=500$ for $L=1$ and $T=1$, with $\mu \in [0.1, 0.15, 0.2, 0.25, 0.3]$, $k=2$, $n_s=8$. The superposed solution $q$ and the individual stationary waves $q^1$ and $q^2$ for all the five different $\mu$ are shown in the three rows of \Cref{fig:synthetic_data} respectively. Note that the pictures have the same structure $Q_{i, j + N_t*p} = q(x_i, t_j, \mu_p)$ as the snapshot matrices $Q, Q^1, Q^2$.
\\
However, for further analysis, we assume that we have access to only $Q$ for every $\mu$ so for $M=500$, $N_t=500$ and $N_p=5$ the collection of snapshots results in a matrix of size $Q\in \mathbb{R}^{M \times N} = \mathbb{R}^{500 \times 2500}$. In the first step, we apply sPOD on this matrix $Q$ and obtain a low-dimensional representation. We let the algorithm run for optimal decomposition such that the $E^{\mathrm{sPOD}} \sim \mathcal{O}(10^{-4})$. We use the ansatz $Q \approx \tilde{Q} = T^{\bm{\Delta}^1}Q^1 + T^{\bm{\Delta}^2}Q^2$. The result of the decomposition is shown for $\mu_0$ in \Cref{fig:synthetic_spod_frames}.
The low-dimensional description for each of the sPOD decomposed frames $Q^k$ for $\mu_0$ is given as:
\begin{equation}\label{eq:sPOD_frame_decomp}
    Q^k = q^k(x, t, \vec{\mu}) \approx \sum^{r_k}_{i=1} a^k_i(t, \vec{\mu})\phi^k_i(x), \quad k=1, 2
\end{equation}
The time amplitude matrix $A^k$ can be extracted as shown in \Cref{eq:sPOD_extract} and also we already have access to the shifts for all $k$. Thus for training the neural network model, we assemble the time amplitude matrix $\hat{A}$ and the parameter matrix $P$ as shown in \Cref{eq:par_ta_matr_sPOD_del} and \Cref{eq:par_matr}. For our problem $\hat{A} \in \mathbb{R}^{18 \times 2500}$ and $P \in \mathbb{R}^{2 \times 2500}$. We run the training loop for $N_{\mathrm{epochs}}=150000$ with batch size $N_b=50$. For testing, we choose $\mu = [0.23]$ and use the trained model to predict the time amplitudes and the shifts which are shown in \Cref{fig:synthetic_time_amplitudes_pred}.   
We assess the prediction accuracy of the proposed methods through \Cref{fig:synthetic_full_err}. In the first plot, we see that with an increasing number of modes the $E^{\mathrm{sPOD}}$ decreases rapidly compared to $E^{\mathrm{POD}}$ which substantiates our claim about using sPOD for basis construction. Here, it needs to be pointed out that $E_{\mathrm{tot}}$ is bounded from below by the basis reconstruction error behavior thus we see that $E^{\mathrm{POD-NN}}_{\mathrm{tot}}$ is always higher than $E^{\mathrm{POD}}$. Same is true for $E^{\mathrm{sPOD-NN}}_{\mathrm{tot}}$. It is also noted that the prediction accuracy of the sPOD-NN is an order of magnitude better than POD-NN. Surprisingly enough for this particular setting, the neural network is not able to learn the POD modes at all. We see that with an increasing $n_{\mathrm{dof}}$ the $E^{\mathrm{POD-NN}}_{\mathrm{tot}}$ remains very near to 1. The performance for the POD-NN could improve if we add more parameter samples to the training data. However, in this test case, the sPOD-I performs the best. It predicts nearly at the limit of $E^{\mathrm{sPOD}}$, without incurring any extra error in the interpolation step. This can be attributed to the fact that the basis functions for the aforementioned problem are polynomials by design and the sPOD-I uses polynomials for interpolation.

\begin{figure}[h]
\centering
\includegraphics[scale=0.45]{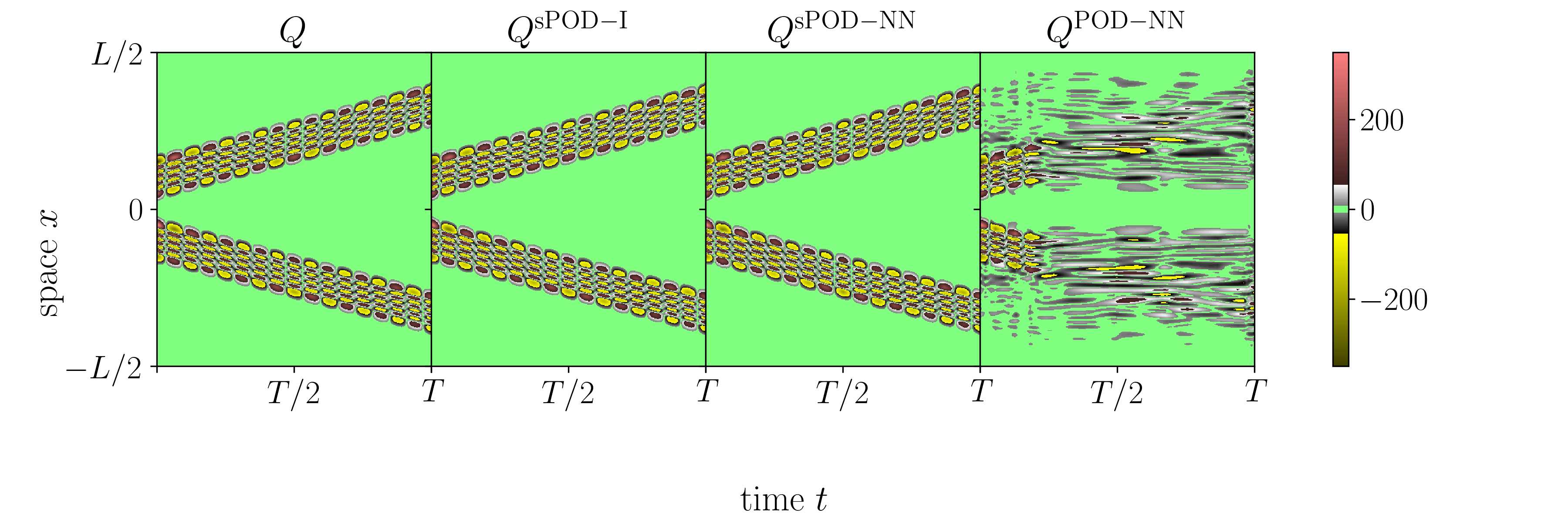}
\caption{Reconstructed snapshot comparison for all the methods}\label{fig:synthetic_snapshot_comp}
\end{figure}

We can get more insight into the prediction accuracy of sPOD-NN by looking at the second plot in \Cref{fig:synthetic_full_err}. It shows the maximum, minimum, and mean of the relative error $E(t, \mu)$ for all the time instances. We see that the maximum error recorded throughout the time frame is around $\sim 0.06$ however, the mean error is always $\sim \mathcal{O}(10^{-3})$.    
We now look at the final reconstructed data in \Cref{fig:synthetic_snapshot_comp}. We observe that the profile with $Q^{\mathrm{POD-NN}}$ is completely distorted, which makes sense when we look at the errors shown in \Cref{fig:synthetic_full_err}. However, there is only a minor visible difference in the profiles of sPOD-NN and sPOD-I compared to the original.   

\end{document}